\documentclass[article,ij4uq]{ij4uq}              

\renewcommand{\dmyy}{17}
\renewcommand{\myyear}{2019}
\renewcommand{\today}{}

\usepackage{amssymb}
\usepackage{latexsym}

\usepackage{amsmath}
\usepackage{dsfont}
\usepackage{comment}

\usepackage{pgfplots}
\usepgfplotslibrary{units}
\usepgfplotslibrary{groupplots}
\usepgfplotslibrary{patchplots}
\usepgfplotslibrary{fillbetween}
\usepackage{pgfplotstable}

\usepackage{pdfpages}
\usepackage[makeroom]{cancel}
\usetikzlibrary{arrows,calc,shapes,arrows,positioning,shadows.blur,shapes.symbols}
\graphicspath{{graphics/}}
\pgfplotsset{compat=newest}
\usepackage{booktabs}

\usepackage{algorithmic} 

\usepackage{glossaries}

\newcommand{\EE}{\mathbb{E}}
\newcommand{\VV}{\mathbb{V}}
\newcommand{\RR}{\mathbb{R}}
\newcommand{\CC}{\mathbb{C}}

\newcommand{\One}{\mathds{I}}
\newcommand{\intd}{\, \text{d}}
\newcommand{\vm}[1]{\ensuremath{\mathbf{#1}}}
\newcommand{\vms}[1]{\ensuremath{\boldsymbol{#1}}}

\newcommand{\veg}[1]{{\boldsymbol{#1}}}
\newcommand{\transpose}{\text{T}}
\newcommand{\eexp}[1]{ \text{e}^{#1} }
\newcommand{\eps}{\varepsilon}

\newcommand{\comments}[1]{}
\DeclareMathOperator*{\argmax}{arg\,max}

\newcommand{\pdf}{\text{pdf}}
\newcommand{\err}{\text{err}}
\newcommand{\up}{\vm p} 
\newcommand{\upj}{p_j} 
\newcommand{\rp}{r}  
\newcommand{\Nmc}{N_{\text{MC}}} 
\newcommand{\Nsc}{N_{\text{SC}}} 
\newcommand{\QoI}{Q} 
\newcommand{\dom}{D} 
\newcommand{\SD}{\Omega_{\text{s}}} 
\newcommand{\IRPo}{T} 
\newcommand{\IRP}{T_{\rp}} 
\newcommand{\dIRP}{T_{\text{d}}} 
\newcommand{\freq}{\omega} 
\newcommand{\scerr}{\tilde{\epsilon}_{\text{sc}}(\up_i, \rp_j)} 
\newcommand{\feerr}{\tilde{\epsilon}_{\text{fe}}(\up_i, \rp_j)} 
\newcommand{\tfeerr}{\epsilon_{\text{fe}}^h(\up_i, \rp_j)} 

\newacronym{fem}{FEM}{finite element method}
\newacronym{fe}{FE}{finite element}
\newacronym{mc}{MC}{Monte Carlo}
\newacronym{sc}{SC}{stochastic collocation}
\newacronym{qoi}{QoI}{quantity of interest}
\newacronym{pec}{PEC}{perfect electric conductor}
\newcommand{\traceop}{\pi_{\text t}}
\newcommand{\Traceop}{\pi_{\text T}}
\newcommand{\epsr}{\epsilon_{\text r}}
\newcommand{\mur}{\mu_{\text r}}

\usepackage[utf8]{inputenc}

\definecolor{tud3d}{RGB}{0, 113, 94}
\definecolor{tud9b}{RGB}{230, 0, 26}

\usepackage{mathtools}
\mathtoolsset{showonlyrefs}

\usepackage{longtable}

\usepackage{pdfcomment}

\newcommand\change[1]{{\color{black}#1}} 

\definecolor{darkpastelgreen}{rgb}{0.01, 0.75, 0.24}
\newcommand\changeAR[1]{{\color{black}#1}} 

\begin{document}

\volume{Volume x, Issue x, \myyear\today}
\title{\change{Yield Optimization based on Adaptive Newton-Monte Carlo and Polynomial Surrogates}}
\titlehead{\change{Yield Optimization based on Adaptive Newton-Monte Carlo and Polynomial Surrogates}}
\authorhead{M. Fuhrländer, N. Georg, U. Römer \& S. Schöps}
\corrauthor[1,2]{Mona Fuhrländer}
\author[2,3]{Niklas Georg}
\author[3]{Ulrich Römer}
\author[1,2]{Sebastian Schöps}
\corremail{fuhrlaender@temf.tu-darmstadt.de}
\corraddress{Institut für Teilchenbeschleunigung und elektromagnetische Felder (TEMF), Technische Universität Darmstadt, Schlossgartenstr. 8, 64289 Darmstadt, Germany}
\address[1]{Institut für Teilchenbeschleunigung und elektromagnetische Felder (TEMF), Technische Universität Darmstadt, Schlossgartenstr. 8, 64289 Darmstadt, Germany}
\address[2]{Centre for Computational Engineering, Technische Universität Darmstadt, Dolivostr. 15, 64293 Darmstadt, Germany}
\address[3]{Institut für Dynamik und Schwingungen, Technische Universität Braunschweig, Schleinitzstr. 20, 38106 Braunschweig, Germany}

\dataO{mm/dd/yyyy}
\dataF{mm/dd/yyyy}

\abstract{
	\change{In this paper we present an algorithm for yield estimation and optimization exploiting Hessian based optimization methods, an adaptive Monte Carlo (MC) strategy, polynomial surrogates and several error indicators.}
	Yield estimation is used to quantify the impact of uncertainty in a manufacturing process. Since computational efficiency is one main issue in uncertainty quantification, we propose a hybrid method, where a large part of a MC sample is evaluated with a surrogate model, and only a small subset of the sample is re-evaluated with a high fidelity finite element model. In order to determine this critical fraction of the sample, an adjoint error indicator is used for both the surrogate error and the finite element error.
	For yield optimization we propose an adaptive Newton-MC method. We reduce computational effort and control the MC error by adaptively increasing the sample size.
	The proposed method minimizes the impact of uncertainty by optimizing the yield.
	It allows to control the finite element error, surrogate error and MC error. At the same time it is much more efficient than standard MC approaches combined with standard Newton algorithms.
}


\maketitle

\section{Introduction}\label{sec:Introduction}

There are many applications where uncertainty quantification and optimization under uncertainty is important. Uncertainty in the manufacturing process may lead to deviations in the design parameters, i.e., geometrical or material parameters, which may lead in turn to rejections due to malfunctioning. In this context, malfunctioning means that pre-defined performance feature specifications are not fulfilled. In order to quantify the impact of uncertainty we define the yield according to~\cite{Graeb_2007aa} 
as the percentage of functioning realizations in a manufacturing process. Thus, yield is mathematically equivalent to the concept of reliability and the relation between yield and failure probability is given in the form \textit{yield = 1 - failure probability}. The topic of yield optimization is motivated by high frequency electromagnetics and circuit design.

In general, it is not possible to carry out yield calculations exactly. Hence, many algorithms have been introduced to this end and the \gls{mc} method is probably the most popular one~\cite{hammersley2013monte}. The main challenge of yield estimation is its high computational cost, since it requires numerous evaluations of the underlying model. In practice, these models are often given by partial differential equations (PDE) of high complexity and can only be solved numerically, with the \gls{fem}, for instance. Since each high fidelity evaluation with \gls{fem} itself may be computationally challenging, a standard \gls{mc} analysis becomes rapidly prohibitive due to limits of computational and / or time resources. 
In this paper we present a hybrid approach for yield estimation combining the efficiency of \change{\gls{sc}} with the accuracy of \gls{mc} for probability estimation. We then present an algorithm for yield maximization, based on a globalized Newton method.

The classical \gls{mc} approach consists in sampling the original high fidelity model, i.e., the highly resolved random \change{\gls{fe}} model. The efficiency of this approach is independent of the number of uncertain parameters and the method does not suffer from the "curse of dimensionality". Still, the sample size required for accurate estimation can be quite large~\cite{Giles_2015aa}. 
There is a lot of research on reducing the computational effort of failure probability or yield estimation. The common goal is to reduce the number of high fidelity evaluations. There are sampling-free methods such as the first order reliability method (FORM) or the second order reliability method (SORM). These methods determine the most probable point, which is the closest point from the parameter domain origin to the  seperating surface between the failure region and the safe region, and employ approximations of the limit state function around this point~\cite{Choi_2012aa, Breitung_1984aa}. 
Investigations in the context of sampling have led to a sample size reduction, e.g., through importance sampling~\cite{Gallimard_2019aa} or subset simulation~\cite{Au_2001aa, Bect_2017aa}. Alternatively or complementarily, the computational effort has been reduced for each sample point, e.g., with surrogate based approaches. In these surrogate methods an approximation (surrogate / response surface) of the original model is built using high fidelity evaluations of a small training set, followed by \gls{mc} sampling of the surrogate model~\cite{Bogoclu_2016aa}. In order to build the surrogate different methods have been employed, e.g., linear regression~\cite{Rao_1999aa}, Gaussian process regression~\cite{Rasmussen_2006aa} or \change{\gls{sc}}~\cite{Babuska_2007aa}. 
\changeAR{In~\cite{Leifsson_2020aa} a combination of two surrogate models, Gaussian process regression and SC is proposed. However,}
%
the accuracy of the surrogate depends on the size of the training set and the number of uncertain parameters. For a large number of uncertain parameters, the computational costs can exceed the costs for \gls{mc}~\cite{Nobile_2008aa}. Furthermore, as shown in~\cite{Li_2010aa}, there are examples where the surrogate model is highly accurate, measured by classical norms or pointwise, but the yield estimator fails drastically. In~\cite{Li_2010aa} a hybrid approach is proposed. Sample points which are close to the limit state function are evaluated based on the high fidelity model, for all remaining sample points the surrogate model is used. Here, the assessment of whether a point is \textit{close} to the interface between failure and safe domain is crucial for the accuracy and the efficiency of the algorithm. To this end, a method using an adjoint error indicator has been presented in~\cite{Butler_2018aa}.
Yield optimization has been carried out in~\cite{Graeb_2007aa}, where a Newton method for optimization was presented, which was combined with the standard \gls{mc} method.

In this paper, we present an algorithm for efficient yield estimation and optimization. For yield estimation we propose a hybrid approach similar to~\cite{Li_2010aa, Butler_2018aa}. Contrary to the approach presented in~\cite{Li_2010aa} we use an adjoint error indicator to identify the aforementioned critical \gls{mc} sample points. Also, contrary to~\cite{Butler_2018aa} we build a polynomial surrogate model based on \change{\gls{sc}}. Furthermore, we consider the \change{\gls{fe}} error in addition to the surrogate error as hybrid distinction criterion. If required, we refine the \change{\gls{fe}} model for a subset of sample points.
We then integrate this hybrid approach into the yield estimation and optimization framework.
The optimization algorithm proposed in this paper is based on a globalized Newton method by~\cite{Ulbrich_2012aa}. For yield estimation, which is necessary in each iteration, we use our previously mentioned hybrid method, and during optimization we adaptively adjust the \gls{mc} sample size. To the best of our knowledge, these are new elements in the context of yield optimization and we call the resulting algorithm adaptive Newton-\gls{mc}. It \change{achieves} an a-priori defined accuracy of the result and significantly reduces computational effort.
Furthermore, we show the applicability of the presented estimation and optimization approaches to problems 
where the performance feature specifications are restrictions involving partial differential equations describing electromagnetic fields, i.e., Maxwell’s equations in frequency domain.

This paper is structured as follows. After setting up the problem in Section~\ref{sec:SetUp}, in Section~\ref{sec:YieldEstimation} we will focus on yield estimation. We briefly review standard \gls{mc} and \change{\gls{sc}}. We then present the hybrid approach combining the two previous ones. In Section~\ref{sec:YieldOptimization} we propose the new adaptive Newton-\gls{mc} method for yield optimization, including the numerical algorithm. Numerical results for the application of electromagnetic field simulation are presented in Section~\ref{sec:NumericalResults} before the paper is concluded in Section~\ref{sec:Conclusion}.

\section{Problem setting}\label{sec:SetUp}
In this paper we consider a PDE with uncertainty in the input data. Details on the differential operator, geometry and boundary conditions will be postponed to a later chapter, which allows us to focus on the main algorithmic aspects for yield estimation and optimization. 
The starting point is the parametric model problem
\begin{equation}
L_{\up,\rp} u_{\rp}(\up) = g_{\rp} \ \ \ \change{\text{in }} D,
\label{eq:Intro_KontMod}
\end{equation}
where $L_{\up,\rp}$ is a linear parametric differential operator, $g$ a forcing term, $\dom \subset \mathbb{R}^d$ a simply connected bounded domain, $\up \in \RR^{{\change{N}_{\up}}}$ the input parameter vector and $r$ the range parameter. The range parameter may refer to frequency or to a temperature for instance, which are not affected by uncertainties.
We assume that the \change{problem is well-posed for all $\up$ and that $\up \mapsto u(\up)$ is a smooth function}, which is often reasonable for parametrized differential equations, see \cite{babuvska2007stochastic} for the case of elliptic problems and \cite{chkifa2015breaking} for other problem classes, for instance. Design objectives are frequently expressed through global quantities, which are modeled in our case as linear functionals of the solution. More precisely, we introduce a \gls{qoi} as 
\begin{equation}
\QoI(\up, \rp) := (q_{\rp}, u_{\rp}(\up))_{D},
\label{eq:QoI_continuous}
\end{equation}
where $q_{\rp} \in L^2(D)$ and $L^2(D)$ denotes the space of complex square-integrable functions with inner product $(\cdot,\cdot)_{D}$.

A \change{\gls{fe}} approach leads to the linear parametric system
\begin{equation}
\vm A_{\up,\rp} \vm u_{\rp}(\up) = \vm f_{\rp},
\label{eq:Intro_DiskrMod}
\end{equation}
where $\vm A_{\up,\rp}\change{\in \mathbb C^{\change{N}_h\times \change{N}_h}}$ denotes the system matrix \change{and $\change{N}_h$ the number of degrees of freedom}. We denote with $u_{h;r}$ the interpolated discrete \change{\gls{fe}} solution, without
explicitly introducing the underlying polynomial \change{\gls{fe}} space. Furthermore, we define the discrete linear \gls{qoi} by
\begin{equation}
\QoI_h(\up, \rp) = \change{ (\vm q_{\rp}, \vm u_{\rp})_{{\mathbb C}^{\change{N}_h}} } 
\label{eq:QoI}
\end{equation}
\change{where $(\cdot, \cdot)_{\mathbb C^{\change{N}_h}}$ refers to the finite-dimensional inner product. }

We assume that the uncertainties originate in the manufacturing process which lead to deviations in the design parameters. These uncertainties are often classified as aleatory. The setting could be generalized by interpreting the computed yield to be conditioned on epistemic uncertainties and by further quantifying these uncertainties as outlined for instance in \cite{papaioannou2019assessment,
	adams2015dakota}. However, since the focus of the present work is on adaptivity and error control in the context of yield estimation, this will not be considered here. The percentage of functioning realizations in mass production is called the yield \cite{Graeb_2007aa}. 
To give a mathematical definition, we model $\up$ as a random  \textit{design parameter} vector, with independent distributed elements $\upj$, $j=1,...,\change{N}_{\up}$. Typically the $\upj$ are assumed to follow a normal distribution, i.e., $\upj \sim \mathcal{N} \left( \overline{\upj}, {\sigma_j} \right)$ with mean value $\overline{\upj} \in \RR$ and standard deviation ${\sigma_j} \in \RR$ and probability density function
\begin{equation}
\pdf_{\mathcal{N}\left( \overline{\upj}, {\sigma_j} \right)}
= \frac{1}{\sqrt{2 \pi {\sigma_j}^2}} \, \eexp{-\frac{(\upj - \overline{\upj})^2}{2 {\sigma_j}^2 }}.
\label{eq:pdf_uni}
\end{equation}
Then, the uncertain parameter $\up$ follows a multivariate normal distribution, i.e., $\up \sim \mathcal{N} \left( \overline{\up}, \vms{\Sigma} \right) $ with mean value $\overline{\up} \in \RR^{\change{N}_{\up}}$ and a diagonal covariance matrix $\vms{\Sigma} \in \RR^{{\change{N}_{\up}} \times {\change{N}_{\up}}}$ and probability density function
\begin{equation}
\pdf_{\mathcal{N}\left( \overline{\up}, \vms{\Sigma} \right)}
= \frac{1}{\change{\left(\sqrt{2 \pi}\right)^{\change{N}_{\up}}} \sqrt{\det \vms{\Sigma}}} \, \eexp{-\frac{1}{2} \left( (\up - \overline{\up})^{\transpose} \vms{\Sigma}^{-1} (\up-\overline{\up}) \right)}.
\label{eq:pdf}
\end{equation}
The normality assumption may be justified by the central limit theorem in the presence of averaging processes or by maximum entropy arguments. Note that, in order to simplify notation, we do not distinguish between a random vector and its realization, whenever there is no confusion in a specific context. Following~\cite{Graeb_2007aa} we further define a \textit{range parameter} \change{$\rp \in \IRP = \left[ \rp_1,\rp_2 \right]$} and the \textit{performance feature specification}
\begin{equation}
\QoI(\up,\rp) \leq c \ \ \forall \rp \in \IRP,
\label{eq:pfs}
\end{equation}
where $c$ is a constant and $\QoI$ the QoI introduced above. 
\change{Note that, without loss of generality, we defined the performance feature specification with an upper bound.} \change{For the sake of notation simplicity, we consider only one. This may be read component-wise, as is usual in optimization.} 
The \textit{safe domain} $\SD$ is the set of all parameters, which fulfill the performance feature specifications, i.e., 
\begin{equation}
\SD := \left\lbrace  \up: \QoI(\up,\rp) \leq c \ \ \forall \rp \in \IRP \right\rbrace.
\label{eq:SafeDomain}
\end{equation}
Then we can express the yield as
\begin{equation}
Y(\overline{\up}) := \EE [\One_{\SD}(\up)]
:= \int_{-\infty}^{\infty} \dots \int_{-\infty}^{\infty} \One_{\SD}(\up) \, \pdf_{\mathcal{N}\left( \overline{\up}, \vms{\Sigma} \right)}(\up)  \intd \up,
\label{eq:Yield}
\end{equation}
where $\EE$ denotes the expected value and $\One_{\SD}(\up)$ the indicator function defined by
\begin{equation}
\One_{\SD}(\up) =
\begin{cases}
1 & \up \in \SD,\\
0 & \, \text{else}.
\end{cases}
\end{equation}
Note that $\overline{\up}$ will be a design parameter during optimization, whereas the covariance is fixed, which is taken into account by our notation in \eqref{eq:Yield}.

\section{Yield Estimation }\label{sec:YieldEstimation}
We proceed by describing a numerical method for yield estimation. The starting point will be a brief description of the MC method, followed by an outline of surrogate modeling based on \change{\gls{sc}}. The section will conclude with a description of a hybrid \change{\gls{mc}} method. 

\subsection{Monte Carlo}
The most straightforward approach in order to estimate the yield, i.e. compute the intergrals of \eqref{eq:Yield}, is a \change{\gls{mc}} analysis \cite{Caflisch_1998aa,hammersley2013monte}. In a \change{\gls{mc}} approach, we consider a large number of independent random variables, distributed in the same way as $\vm p$. The set $\{\up_i\}_{i=1}^{\Nmc}$, where each $\up_i$ represents a realization of the corresponding random variable, is called a sample and $\Nmc$ represents the sample size. At each sample point $\up_i$, we evaluate the high fidelity \gls{fe} model and count the sample points, which fulfill our performance feature specifications. Then we obtain a yield estimator as
\begin{equation}
Y(\overline{\up}) \approx \change{Y_{\text{MC}}(\overline{\up})} := \frac{\# \text{ sample points in } \SD}{\text{ sample size}},
\label{YES_MC_words}
\end{equation}
or equivalently
\begin{equation}
\change{Y_{\text{MC}}(\overline{\up})} = \frac{1}{\Nmc} \sum_{i=1}^{\Nmc} \One_{\SD}(\up_i).
\label{eq:YES_MC}
\end{equation}
MC estimation is based on the law of large numbers, which ensures convergence for $\Nmc \rightarrow \infty$ under mild regularity assumptions on the integrand. Since in practice, the sample size is always finite, we need to estimate the associated error. To this end, we use an error indicator from \cite{Giles_2015aa}. 
An estimator of the approximated yield variance is derived as follows. 
\change{Since} all observations are independent, we obtain
\begin{align}
\VV \left[ \change{Y_{\text{MC}}(\overline{\up})} \right] &= \frac{1}{\Nmc^2} \VV \left[ \sum_{i=1}^{\Nmc} \One_{\SD}(\change{\up_i}) \right] \\
&= \frac{1}{\Nmc^2}  \sum_{i=1}^{\Nmc} \VV \left[\One_{\SD}(\change{\up_i}) \right] \\
&= \frac{1}{\Nmc^2} \Nmc Y(\overline{\up}) (1 - Y(\overline{\up})) \\
&= \frac{Y(\overline{\up})(1-Y(\overline{\up}))}{\Nmc},
\label{eq:VarMC}
\end{align}
where the expectation and variance are now defined with respect to the i.i.d. observations. Then, we derive the standard deviation of the yield estimator as
\begin{equation}
\sigma_Y = \sqrt{\frac{Y(\overline{\up})(1-Y(\overline{\up}))}{\Nmc}} \leq \frac{0.5}{\sqrt{\Nmc}}.
\label{eq:StdMC}
\end{equation}
The standard deviation depends on the size of the yield. For a yield of $50\,\%$ it is maximum and so we obtain the upper bound for the standard deviation given in~\eqref{eq:StdMC}.
Since, the \change{\gls{mc}} estimator is unbiased, the variance is equal to the mean-square error. 
In view of \eqref{eq:StdMC}, this approach guarantees a high accuracy for a large sample size, but it converges slowly with $\mathcal{O} \left(1 / \sqrt{\Nmc} \right)$. In many cases this is unaffordable due to the large number of expensive function evaluations required~\cite{Giles_2015aa}.

\subsection{Stochastic Collocation and Error Estimation}\label{subsubsec:SC}
To reduce the computational complexity of sampling the underlying \gls{fe} solver, surrogate models can be employed. Based on the assumption that \change{ a map $X: \RR^{{\change{N}_{\up}}} \times T_r \rightarrow \mathbb C$ (where $X$ might refer to the \gls{qoi} $\QoI_h$, for instance)} is well-defined 
and sufficiently smooth, \change{we denote by $\tilde{X}$ the surrogate approximation defined by}
\begin{equation}
\tilde{X} \left( \up, \rp \right) = \sum_{\change{i=1}}^{\change{\Nsc}} \alpha_{i} \left( \rp \right) \Phi_i \left( \up \right),
\label{eq:ApproxPoly}
\end{equation}
where \change{$\Nsc$ is the number of interpolation nodes,} $\Phi_i:\RR^{{\change{N}_{\up}}} \rightarrow \mathbb R$ are multivariate global polynomial basis functions with respect to $\up$ and $\alpha_i:T_r\rightarrow\mathbb C$ denote the corresponding coefficients. Such a construction is appealing, as spectral convergence with respect to the polynomial degree can be expected \cite{xiu2002}. In this work, we compute such approximations based on the \change{\gls{sc}} method \cite{xiu2005high,babuvska2007stochastic}. In particular, the surrogate model is obtained by evaluating \eqref{eq:Intro_DiskrMod} for a set of multivariate interpolation nodes $\{\up^{(i)}\}_{i=0}^{\change{\Nsc-1}}$ and enforcing the corresponding collocation conditions on the surrogate model. The choice of the multivariate nodes $\up^{(i)}$ is crucial for the efficiency of \change{\gls{sc}}. To this end, we first consider the tensor grid of univariate interpolation nodes $\{p_1^{(i)}\}_i \times \{p_2^{(i)}\}_i \times \ldots \times \{p_M^{(i)}\}_i$. Employing all points of the grid is computationally intractable for many parameters. Sparse-grids \cite{bungartz2004} are a viable alternative, where a subset of points, which do not significantly contribute to the approximation accuracy is neglected. In this work, we use an algorithm proposed in \cite[Algorithm 2]{georg2018}, which constructs the sparse-grid adaptively. For convenience of the reader, we recall the main ideas in the following. 

The algorithm is based on weighted Leja nodes \cite{narayan2014} which are defined recursively by an optimization problem, i.e., univariate weighted Leja nodes  $\{p_m^{(i)}\}_i \subset \mathbb R$ are obtained as 
\begin{equation}
p_m^{(I)} = \argmax_{p_m \in \mathbb R} \sqrt{w(p_m)}\prod_{i=0}^{I-1} |p_m-p_m^{(i)}|,
\end{equation}
where the weight function $w(p_m)$ is typically chosen as the probability distribution of the corresponding input parameter, i.e., $w(p_m)= \pdf_{\mathcal{N}\left( \overline{p_m}, {\sigma_m} \right)}$, and for the first node we set $p_m^{(0)}=0$. Leja nodes are well suited for adaptive approximations in higher dimensions, since they are, by construction, nested and allow for a granular refinement \cite{narayan2014}. To steer the adaptive selection of the corresponding multivariate nodes, an adjoint error indicator \cite{butler2012posteriori, butler2013propagation} is employed. To this end, we introduce the dual problem to \eqref{eq:Intro_DiskrMod}, which is given by
\begin{equation}
\vm A^{\star}_{\up,\rp} \, \vm z_{\rp}(\up) = \vm q_{\rp},
\end{equation}
where $\vm A^{\star}$ denotes the Hermitian transpose of $\vm A$. In addition to \change{the polynomial approximation of the \gls{qoi} $\tilde Q_h$}, we construct polynomial approximations of the mappings $\vm u, \vm z:\mathbb R^{\change{N}_{\up}}\times T_\text{r}\rightarrow \mathbb C$, where the same collocation points as for the \gls{qoi} are employed, cf. \cite{jakeman2015}. The resulting approximations are denoted as $\tilde{\vm u}, \tilde{\vm z}$. We are then interested in the error 
\begin{align}
\epsilon_\text{sc}(\up, r) &= \change{ \left| \QoI_h(\up,r)-\tilde \QoI_h(\up,r) \right|} \\
& = \change{\left| \bigl(\vm q_{\rp}, \vm u_{\rp}(\up)\bigr)_{{\mathbb C}^{\change{N}_h}} - \bigl(\vm q_{\rp}, \tilde{\vm u}_{\rp}(\up)\bigr)_{{\mathbb C}^{\change{N}_h}} \right|}\\
& = \change{\left|\bigl(\vm z_{\rp}(\up), \vm A_{\up,\rp} \, \vm u_{\rp}(\up)\bigr)_{{\mathbb C}^{\change{N}_h}} - \bigl(\vm z_{\rp}(\up), \vm A_{\up,\rp} \, \tilde{\vm u}_{\rp}(\up)\bigr)_{{\mathbb C}^{\change{N}_h}} \right|}\\
&= \change{\left| \bigl(\vm z_{\rp}(\up),  \vm f_{\rp} - \vm A_{\up,\rp} \, \tilde{\vm u}_{\rp}(\up) \bigr)_{{\mathbb C}^{\change{N}_h}} \right|}. \label{eq:parametric_error}
\end{align}
The evaluation of \eqref{eq:parametric_error} would always require the computation of $\vm z$, i.e., the solution of the high fidelity adjoint problem. Hence, following \cite{jakeman2015}, we employ the error indicator 
\begin{equation}
\tilde{\epsilon}_\mathrm{sc}(\up, r) := \change{\left | \bigl(\tilde{\vm z}_{\rp}(\up), \vm f_{\rp} - \vm A_{\up,\rp} \, \tilde{\vm u}_{\rp}(\up) \bigr)_{{\mathbb C}^{\change{N}_h}} \right|}. \label{eq:parametric_error_indicator}
\end{equation}
It should be noted that, under mild assumptions, cf. \cite{butler2012posteriori, georg2018}, the error ocurring when $\vm z$ is replaced with $\tilde{\vm z}$ is of higher order. The error indicator is then used to select interpolation nodes which are admissible for refinement of the approximations until a given computational budget is reached and the algorithm terminates. For further details on the employed adaptive sparse-grid interpolation scheme, we refer to \cite{georg2018}. 
Once an accurate surrogate model is available, it can then be used as an inexpensive substitute of \eqref{eq:QoI} for an extensive \gls{mc} analysis \eqref{eq:YES_MC}.

Adjoint techniques can further be used to \change{estimate} the \change{\gls{fe}} error following \cite{becker2001optimal,eriksson1995introduction}.  However, in this case, the continuous adjoint equation is required, which reads
\begin{equation}
L_{\up,\rp}^* z_{\rp}(\up) = q_r \ \ \ \change{\text{in }} D,
\label{eq:Intro_KontModAdj}
\end{equation}
where $L_{\up,\rp}^*$ denotes the adjoint operator with respect to the inner product $(\cdot,\cdot)_D$. With this notation at hand, we derive the following identity for the \gls{fe} error
\begin{align}
\epsilon_\text{fe}(\up, r) &= \change{\left| (q_r,u_r(\up) - u_{h;r}(\up))_D \right|} \\
&= \change{\left|(L_{\up,\rp}^* z_{\rp}(\up),u_r(\up) - u_{h;r}(\up))_D \right|}\\
&= \change{\left|(z_{\rp}(\up),L_{\up,\rp} (u_r(\up) - u_{h;r}(\up)))_D \right|}\\
&= \change{\left|(z_{\rp}(\up),g_\rp - L_{\up,\rp} u_{h;r}(\up))_D \right|}.
\label{eq:fe_error}
\end{align}
A computable expression can only be obtained if the adjoint is replaced with a \change{\gls{fe}} approximation. However, we cannot simply employ $z_{h;\rp}$ as it is orthogonal to the residual. Hence, a higher order adjoint is required for the \gls{fe} error, contrary to the surrogate error \eqref{eq:parametric_error_indicator}. A discussion can be found in \cite{butler2012posteriori}. Hence, we approximate the adjoint solution on a refined grid, but other options, such as higher polynomial degrees or recovery techniques \cite{zienkiewicz1992superconvergent}, are equally applicable. 

Finally, an error identity comprising both SC and \gls{fe}-contribution is obtained as 
\begin{align}
\change{\left| Q(\up,r) - \tilde{Q}_h(\up,r) \right|} &\change{ \leq \Big| Q(\up,r) - Q_h(\up,r) \Big| + \left| Q_h(\up,r) - \tilde{Q}_h(\up,r) \right|} \\
& \change{\approx \left| \bigl(z_{h/2;r}(\up),g_\rp - L_{\up,\rp} u_{h;r}(\up)\bigr)_D \right|  + \left| \bigl(\tilde{\vm z}_{\rp}(\up), \vm f_{\rp} - \vm A_{\up,\rp} \, \tilde{\vm u}_{\rp}(\up) \bigr)_{{\mathbb C}^{\change{N}_h}} \right|}.
\label{eq:combined-error}
\end{align}
\change{
	The second term is immediately identified as $\tilde{\epsilon}_\mathrm{sc}(\up, \rp)$, which uses the surrogate approximations $\tilde{\vm z}_r(\up)$ and $\tilde{\vm u}_r(\up)$, and can therefore easily be evaluated for all $\up$. However, the first term is identified as a computable approximation to $\epsilon_\mathrm{fe}(\up,\rp)$, which we use, along with \eqref{eq:ApproxPoly}, to build the surrogate approximation $\tilde{\epsilon}_\mathrm{fe}(\up, \rp)$.} 
\change{The separation of the FE error and the SC error by the triangle inequality is a rather conservative choice to define the total error. We come back to this point at a later stage.}
We note, that the combined estimation of deterministic and stochastic discretization errors, has for example also been considered in \cite{butler2013propagation}, in the context of the stochastic Galerkin method for time-dependent forward and inverse problems.

\subsection{Hybrid approach}\label{subsubsec:Hybrid}

The number of collocation points $N$, for which the high fidelity \gls{fe} model needs to be solved, depends on the number of uncertain parameters and the polynomial degree the surrogate model is supposed to have. This number grows rapidly with the number of parameters (''curse of dimensionality'')~\cite{bellman1961curse}. For adaptive sparse grids the required \gls{fe} solver calls can be reduced significantly. However, we know from~\cite{Li_2010aa} that yield estimation may produce erroneous results even though the surrogate model may be highly accurate.

The aim of the hybrid approach is to restore the accuracy of the \gls{mc} method while relying on surrogate modeling as much as possible to enhance the numerical efficiency. We propose a particular hybrid approach, which is an extension of the one presented in~\cite{Li_2010aa}. The main difference lies in the selection of sample points which have \change{to} be re-evaluated with the high fidelity model. These points are referred to as critical sample points in the following. In~\cite{Li_2010aa} a tube around the boundary of the failure domain is defined, where the tube size is either fixed in advance, or determined iteratively by an algorithm which adds critical samples points until some error bound is satisfied. 
In comparison to~\cite{Butler_2018aa} the method we propose is
using \change{\gls{sc}} with Leja nodes as surrogate model (see Section~\ref{subsubsec:SC}). Also, in addition to the surrogate model error (SC error), we also consider the \change{\gls{fe}} error in order to determine the critical sample points. Both error contributions are estimated by the adjoint error indicator, according to~\eqref{eq:combined-error}.

\begin{figure}[t]
	\begin{center}
		\includegraphics[]{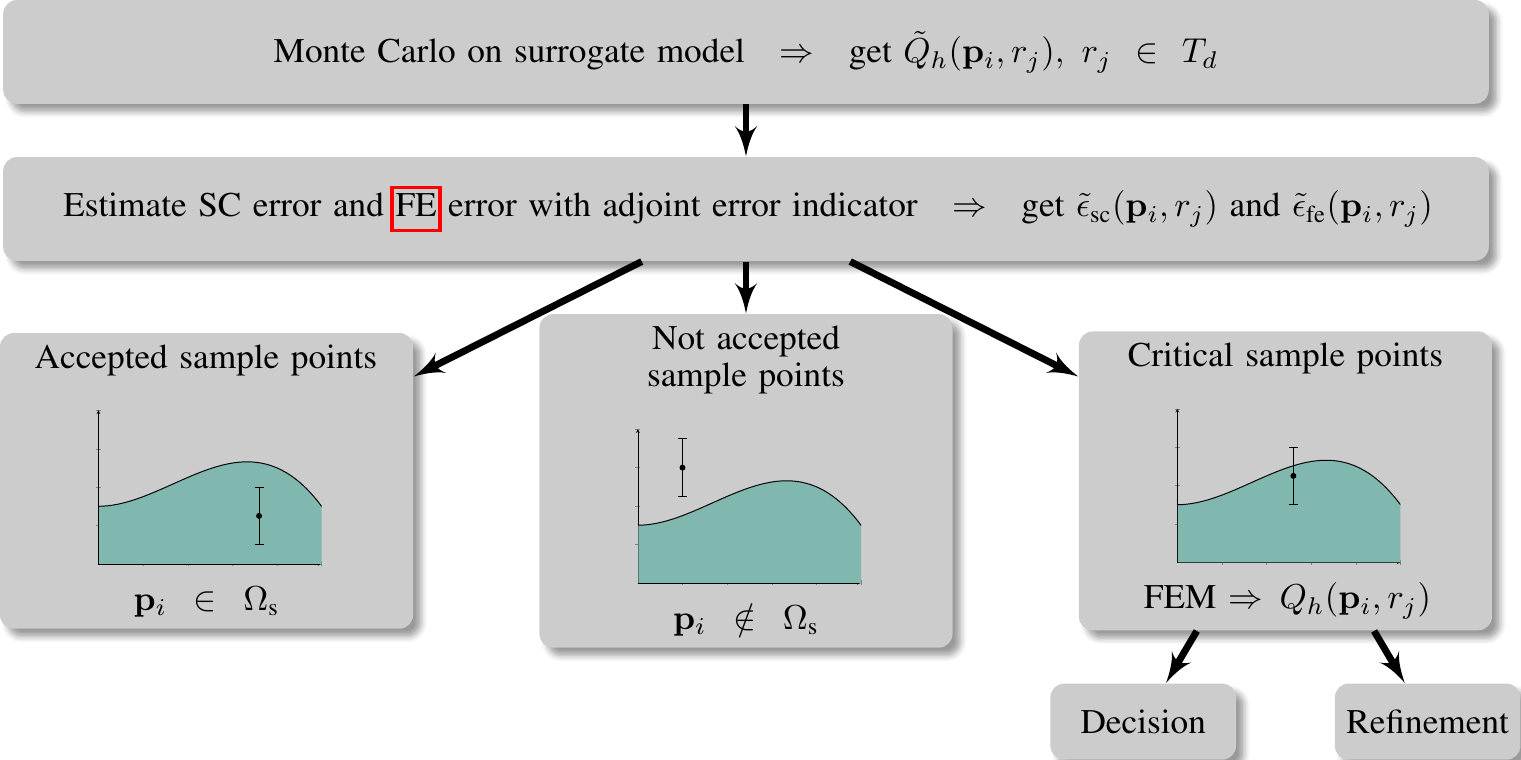}
		\caption{Scheme of the hybrid approach.}
		\label{fig:Hybrid_scheme}
	\end{center}
\end{figure}
\changeAR{In the following we assume for simplicity of notation the \gls{qoi} to be real valued.}
%
Our procedure is summarized in Figure~\ref{fig:Hybrid_scheme}.
The first step is to build a surrogate model and to carry out a \gls{mc} analysis with it.
Then, we use an adjoint error indicator to quantify both the \gls{fe} and surrogate error as
\begin{equation}
\scerr \text{ and } \feerr \ \ \forall i=1,\dots, \Nmc, \ \forall j = 1,\dots,\left| \dIRP \right|,
\label{eq:errors_in_est}
\end{equation}
where $\dIRP$ is a discrete subset of $\IRP$.
We then verify whether the approximated \gls{qoi} value, taking into account the aforementioned errors, meets the requirements. To this end, we define the interval
\begin{equation}
\change{\mathcal{I}^1_{\epsilon}(\up_i,\rp_j) = \left[  \tilde{\QoI}_h(\up_i,\rp_j)  - s  \left(  \scerr  +  \feerr \right),  \tilde{\QoI}_h(\up_i,\rp_j)  + s  \left( \scerr + \feerr  \right) \right]},
\end{equation}
where $s \geq 1$ indicates a safety factor.
If the performance feature specifications are fulfilled (or not fulfilled) for the whole interval $\change{\mathcal{I}^1_{\epsilon}}$, we can classify the sample point $\up_i$ as accepted (or not accepted). If the performance feature specifications are fulfilled only for a subset of the interval $\change{\mathcal{I}^1_{\epsilon}}$, we classify the sample point as critical.

For all critical sample points the high fidelity \gls{fe} model will be evaluated, hence, we obtain $\QoI_h(\up_i,\rp_j)$. For these points, the surrogate error is zero, however, the \gls{fe} error remains unchanged. The new interval we have to examine is given by
\begin{equation}
\change{\mathcal{I}^2_{\epsilon}(\up_i,\rp_j) = \left[  \QoI_h(\up_i,\rp_j)  - s  \left( 0 +  \feerr  \right), \QoI_h(\up_i,\rp_j)  + s\left( 0 +  \feerr  \right) \right]}.
\end{equation}
Applying the same rules as above, the sample points are again classified either as \textit{accepted} or \textit{not accepted}. If the sample point is not identified as critical, we continue with the next sample point. Else, we refine the mesh of the \gls{fe} model and re-evaluate $\QoI_h(\up_i,\rp_j)$ and the \gls{fe} error $\tfeerr$. We continue this procedure until the sample point is not critical anymore or a maximal number of refinement steps is reached. 
\change{In this manner we obtain an accuracy comparable to the pure \gls{mc} approach, using the finest refinement. The only difference would occur for sample points whose errors were greatly underestimated with the adjoint error indicators and which were therefore wrongly accepted or rejected instead of being classified as critical sample points.}
The decision process for one sample point $\up_i$ in one range parameter point $\rp_j$ is reported in Algorithm~\ref{algo:HybridDecision}. 
\begin{algorithm}[!t] 
	\caption{Hybrid decision}
	\begin{algorithmic}[1]
		\STATE{\textbf{Input:} sample point $\up_i$, range parameter point $\rp_j$, safety factor $s$}
		\STATE{Evaluate surrogate model and set \\
			~~~~$\QoI = \change{ \tilde{\QoI}_h(\up_i,\rp_j) }$\\
			~~~~$\epsilon = \change{ \scerr  +  \feerr }$}
		\WHILE{max. refinement not reached}
		\IF{$\QoI-s \, \epsilon > c$}
		\STATE{classify $\up_i$ as \textit{not accepted}, i.e., $\up_i \notin \SD$ (middle picture in Fig. \ref{fig:Hybrid_scheme})\\
			continue with next sample point $\up_{i+1}$}
		\ELSIF{$\QoI+s \, \epsilon \leq c$}
		\STATE{sample point $\up_i$ accepted for this range parameter point $\rp_j$
			\IF{all $\rp_j$ checked}
			\STATE{classify $\up_i$ as \textit{accepted} , i.e., $\up_i \in \SD$ (left picture in Fig. \ref{fig:Hybrid_scheme})\\
				continue with next sample point $\up_{i+1}$}
			\ELSE
			\STATE{check next range parameter point $\rp_{j+1}$}
			\ENDIF
		}
		\ELSE
		\STATE{sample point $\up_i$ is \textit{critical}
			\IF{first loop}
			\STATE{Evaluate \gls{fe} model and set \\
				~~~~$\QoI = \change{ \QoI_h(\up_i,\rp_j) }$\\
				~~~~$\epsilon = \change{ \feerr }$}
			\ELSE
			\STATE{Refine the mesh with $h=h/2$\\
				Evaluate \gls{fe} model and set \\
				~~~~$\QoI = \change{ \QoI_h(\up_i,\rp_j) }$\\
				~~~~$\epsilon = \change{ \tfeerr }$}
			\ENDIF
		}
		\ENDIF
		\ENDWHILE
		\IF{sample point $\up_i$ still \textit{critical} with last refinement}
		\STATE{classify $\up_i$ according to $\QoI$ with the finest mesh into \textit{accepted} or \textit{not accepted}}
		\ENDIF
	\end{algorithmic}
	\label{algo:HybridDecision}
\end{algorithm}

\change{
	The following paragraph is dedicated to the choice of the safety factor $s$. The FE error indicator $\feerr$ and the SC error indicator $\scerr$ defined in Section~\ref{subsubsec:SC} are not strict upper bounds. Therefore, we introduce the safety factor. 
	To determine the size of the safety factor, we generate a small random sample and evaluate it on the surrogate model $\tilde{\QoI}_h(\up_i,\rp_j)$ and on the original model $\QoI_{h}(\up_i,\rp_j)$ (with the finest mesh examined). Next we consider the maximum of the ratios of $\feerr + \scerr$ and $|\QoI_h(\up_i,\rp_j)-\tilde{\QoI}_h(\up_i,\rp_j)|$ to derive the safety factor. 
	As with the computation of the total error~\eqref{eq:combined-error}, we choose the safety factor rather conservatively. 
	This may result in too many sample points being classified as critical, thus increasing the computational effort of the hybrid approach. However, it avoids the misclassification of sample points and thus leads to a higher accuracy.
	Here, the safety factor has been set to $s=2$.
}

The performance feature specifications have to be fulfilled for all $\rp \in \IRP$, or at least for all test range parameter points $\rp_j \in \dIRP$. Thus, 
if one sample point $\up_i$ fulfills the requirements for a specific range parameter point, the test needs to be carried out for the remaining range parameter points as well.
However, if $\up_i$ fails to fulfill the requirements for a single arbitrary range parameter point, it is immediately classified as \textit{not accepted}.
Thereby, we can avoid the computational effort of evaluating the remaining range parameter points. This strategy is also applied for the standard \gls{mc} method and the \change{\gls{sc}} surrogate-based \gls{mc} method. 
\change{In the hybrid method we can further benefit from the fact, that we know the \change{\gls{sc}} results for the \gls{qoi}. As the performance feature specification is defined as an upper bound, we assume that for larger values of $\tilde{\QoI}_h(\up_i,\rp_j)$ it is more likely that $Q_h(\up_i,\rp_j)$ does not fulfill the requirements. Hence, we order the range parameter points according to $\tilde{\QoI}_h(\up_i,\rp_j)$ and start examining the range parameter point satisfying
	\begin{equation}
	\argmax_{\rp_j \in \dIRP} \tilde{\QoI}_h(\up_i,\rp_j).
	\end{equation}}

In total, three different errors have to be considered within the yield estimation process: the \gls{mc} error, the \gls{fe} error and the error of the surrogate model, in our case the SC error. The hybrid approach proposed in this paper takes into account the surrogate and \gls{fe} error. The \gls{fe} error depends on the refinement of the mesh. Instead of evaluating the entire \gls{mc} sample (or all critical sample points in a hybrid approach) with the finest mesh, we start with a coarse mesh, calculate the error indicator and refine the mesh if necessary. Thereby, the \gls{fe} error is controlled and reduced if required and unnecessary computational effort avoided.
The SC error is controlled by calculating an adjoint error indicator after building the surrogate model. If the sum of both indicators is too large, a sample point may be classified as critical. In this case, we evaluate the \gls{fe} model and the associated SC error vanishes.
In order to control the \gls{mc} error, we define a target accuracy by a maximum value of the standard deviation $\sigma_Y$ and determine the minimum sample size needed by~\eqref{eq:StdMC}.

\section{Yield Optimization}\label{sec:YieldOptimization}

\subsection{General Newton approach}\label{subsec:GeneralNewton}

The idea of yield optimization is to change the mean value of the uncertain parameter, i.e., $\overline{\up}$, in order to maximize the yield. We can formulate the optimization problem as follows
\begin{equation}
\max_{\overline{\up}}Y(\overline{\up}) =\max_{\overline{\up}}
\int_{-\infty}^{\infty} \dots \int_{-\infty}^{\infty} \One_{\SD}(\up) \, \pdf_{\mathcal{N}\left( \overline{\up}, \vms{\Sigma} \right)}(\up)  \intd \up.
\label{eq:OptProb}
\end{equation}
Let the uncertain parameter $\up$ be modeled as a normally distributed random variable. Then, since only the probability density function of the uncertain parameter $\up$ depends on the optimization variable $\overline{\up}$, from \eqref{eq:Yield} we can derive the gradient and the Hessian of the yield according to~\cite{Graeb_2007aa}. \change{To this end, we first introduce the mean and covariance of $\up$, conditional to the event $\up \in \SD$, given as}
\begin{align}
\overline{\up}_{\SD} &= \EE_{\pdf_{\SD}} \left[\up \right] 
= \int_{-\infty}^{\infty} \dots \int_{-\infty}^{\infty} \up \ \change{ \pdf_{\SD}(\up)} \intd \up,
\label{eq:MWAbV}\\
\vms \Sigma_{\SD} &= \EE_{\pdf_{\SD}}  \left[(\up - \change{\overline{\up}_{\SD}}) \ (\up - \change{\overline{\up}_{\SD}})^{\transpose}\right] \notag \\
&=  \int_{-\infty}^{\infty} \dots \int_{-\infty}^{\infty} (\up - \overline{\up}_{\SD}) \ (\up - \overline{\up}_{\SD})^{\transpose} \ \change{\pdf_{\SD}(\up)} \intd \up \label{eq:KovAbV}
\end{align}
\change{where}
\begin{equation}
\pdf_{\SD}(\up) = \frac{1}{Y(\overline{\up})} \, \One_{\SD}(\up) \, \pdf_{\mathcal{N}(\overline{\up}, \vms \Sigma)}(\up).
\end{equation}
\change{These conditional moments} can be estimated by
\begin{align*}
\hat{\overline{\up}}_{\SD} &= \frac{1}{\change{N_{\SD}}} \sum_{i=1}^{\Nmc} \One_{\SD}(\up_i) \ \up_i,\\
\hat{\vms \Sigma}_{\SD} &= \frac{1}{\change{N_{\SD}}-1} \sum_{i=1}^{\Nmc} \One_{\SD}(\up_i) \ \left(\up_i - \change{\hat{\overline{\up}}_{\SD}}) \ (\up_i - \change{\hat{\overline{\up}}_{\SD}}\right)^{\transpose},
\end{align*}
where $\up_i, i=1,...,\Nmc$ are independent observations of the random variable $\up$ and $\change{N_{\SD}}$ indicates the number of sample points within the safe domain.
Using these formulations, the gradient and the Hessian of the yield with respect to $\overline{\up}$ can be written as
\begin{align}
\nabla_{\overline{\up}} Y(\overline{\up}) &=  
\int_{-\infty}^{\infty} \dots \int_{-\infty}^{\infty} \One_{\SD}(\up) \, \nabla_{\overline{\up}} \pdf_{\mathcal{N}\left( \overline{\up}, \vms{\Sigma} \right)}(\up)  \intd \up
\approx Y(\overline{\up}) \, \vms \Sigma^{-1} \, (\overline{\up}_{\SD}-\overline{\up})
\label{eq:Grad}\\
\nabla_{\overline{\up}}^2 Y(\overline{\up}) &= 
\int_{-\infty}^{\infty} \dots \int_{-\infty}^{\infty} \One_{\SD}(\up) \, \nabla_{\overline{\up}}^2 \pdf_{\mathcal{N}\left( \overline{\up}, \vms{\Sigma} \right)}(\up)  \intd \up \\
&\approx Y(\overline{\up}) \, \veg \Sigma^{-1} \, 
\left( {\vms \Sigma}_{\SD}+(\overline{\up}_{\SD}-\overline{\up}) \,
(\overline{\up}_{\SD}-\overline{\up})^{\text{T}} - \vms \Sigma \right) \,
\vms \Sigma^{-1}.
\label{eq:Hess}
\end{align}
A detailed derivation can be found in~\cite{Graeb_2007aa}. It should be mentioned that we first differentiate and then discretize. Hence, this gradient does not necessarily coincide with the gradient obtained by differentiating after discretization.

\begin{algorithm}[htb] 
	\caption{Globalized Newton method}
	\begin{algorithmic}[1]
		\STATE \textbf{Input:} Starting point $\overline{\up}^0 \in \RR^{\change{N}_{\up}}$, $\beta \in (0,1)$, $\gamma \in (0,1)$, $\change{\upvarphi}_1, \change{\upvarphi}_2 > 0$, $q>0$
		\STATE \textbf{Output:} Optimal solution $\overline{\up}^{\star}$
		\WHILE{ $\nabla Y\left(\overline{\up}^k\right) \neq 0$  \AND $\left\| \overline{\up}^k -  \overline{\up}^{k-1} \right\| > 0 $}
		\STATE {Calculate $\vm d^k$ by solving Newton's equation $\nabla^2  Y\left(\overline{\up}^k\right)\vm d^k = -\nabla  Y\left(\overline{\up}^k\right)$.}
		\IF {"Calculation of $\vm d^k$ possible" \textbf{and} $-\nabla  Y\left(\overline{\up}^k\right)\vm d^k \geq \min \left( \change{\upvarphi}_1, \change{\upvarphi}_2 \left\| \vm d^k \right\|^q \right) 										\left\| \vm d^k \right\|^2$}
		\STATE {Set search direction $\vm s^k = \vm d^k$.}
		\ELSE
		\STATE {Set search direction $\vm s^k = -\nabla  Y\left(\overline{\up}^k\right)$.}
		\ENDIF
		\STATE {Determine step size with Armijo rule, i.e., search for largest $\sigma^k \in \left\{ \beta^0, \beta^1, \beta^2,... \right\}$ \\
			~~~~such that: $Y\left(\overline{\up}^k +\sigma^k \vm s^k  \right) - Y\left(\overline{\up}^k\right) \leq \sigma^k \gamma \nabla  Y\left(\overline{\up}^k\right)^{\transpose} \vm s^k$. } 
		\STATE {Set $\overline{\up}^{k+1}=\overline{\up}^k + \sigma^k \vm s^k$ and $k=k+1$.}
		\ENDWHILE
	\end{algorithmic}
	\label{algo:globNewton}
\end{algorithm}
The fact that we have given the gradient and the Hessian in analytical form allows us to use a \change{Hessian} based optimization algorithm, such as the globalized Newton method~\cite{Ulbrich_2012aa} as proposed in~\cite{Graeb_2007aa}. A pseudo code is given in Algorithm~\ref{algo:globNewton}.
The associated parameters have been set as follows
\begin{equation}
\beta = \frac{1}{2},~~ \gamma = \frac{1}{100},~~ \change{\upvarphi}_1=\change{\upvarphi}_2=10^{-6},~~ q = \frac{1}{10}.
\label{eq:ParameterGlobNewton}
\end{equation}

In this paper we assume that all uncertain parameters are optimization variables and vice versa. Little modifications in the algorithm also cover other cases. 
\changeAR{Additional deterministic optimization variables would appear in the indicator function. Thus, the analytical formulation of the gradient (and the Hessian) does not hold. Instead a finite difference approximation can be used or a negligible uncertainty (noise) can be assigned.}
%
If, instead, there are uncertain parameters $\vm u$, which are not optimization variables, they have to be considered during yield estimation, which can be achieved by setting $\up' = \left[ \up, \vm u \right]^{\transpose}$. Nevertheless, during optimization we only use $\up$, e.g. to calculate $\vms \Sigma$, $\overline{\up}_{\SD}$, $\vms \Sigma_{\SD}$, etc.

If additional optimization variables without uncertainty are present, these additional optimization variables appear in the indicator function, since the safe domain depends on them.
Thus, the analytical formulation of the gradient (and the Hessian) does not hold anymore. 
Instead, for the partial deviations regarding the optimization variables without uncertainty, finite differences can be used.
Another option constists in avoiding the singularity of the covariance matrix by adding a negligible small uncertainty (noise) to the not uncertain variables.

\subsection{Adaptive Newton-\gls{mc}}\label{subsec:AdaptNewtonMC}

The size of the \gls{mc} sample is crucial, not only for accuracy but also for the efficiency of the algorithm. According to~\eqref{eq:StdMC}, for yield estimation we can use the \gls{mc} error indicator to determine the sample size depending on the desired accuracy. For yield optimization, the situation is more involved. The accuracy of yield estimators at intermediate steps of the Newton algorithm is not essential to obtain a satisfying final result. In each individual iteration, it is sufficient to obtain a gradient that indicates the right direction. 
The stochastic gradient approach also deals with approximated or inexact gradients, used during the optimization process, see~\cite{Geiersbach_2019aa} for example. However, our approach uses more sample points than usual in the stochastic gradient approach, but we also calculate the objective function with the reduced sample.
Only towards the termination of the algorithm, a very accurate gradient may be decisive to accurately determine the optimal solution. Our algorithmic construction ensures that the high, pre-defined, accuracy requirements at the final stages of the algorithm are fullfilled. More precisely, we propose the following adaptive Newton-\gls{mc} approach. The optimization method is based on a globalized Newton method, as described in Algorithm~\ref{algo:globNewton}. We start with a very small sample size and proceed with a few \textit{fast} initial Newton iterations. If no further yield improvement is observed during the iteration process, the globalized Newton method described in Algorithm~\ref{algo:globNewton} would stop. Here, instead, we increase the number of \gls{mc} observations until an improved yield is observed or a target accuracy is reached, then we start the next Newton iteration. Only when the target accuracy has been reached and the yield is not improving anymore, the algorithm terminates.

\begin{algorithm}[htb] 
	\caption{Adaptive Newton-MC}
	\begin{algorithmic}[1]
		\STATE \textbf{Input:} Starting point $\overline{\up}^0 \in \RR^{\change{N_{\up}}}$, max. std. $\sigma_{Y,\mathrm{max}}$, starting sample size $\Nmc^{\text{start}}$, $\beta \in (0,1)$, $\gamma \in (0,1)$, $\change{\upvarphi}_1, \change{\upvarphi}_2 > 0$, $q>0$
		\STATE \textbf{Output:} Optimal solution $\overline{\up}^{\star}$
		\WHILE{ $\nabla Y\left(\overline{\up}^k\right) \neq 0$  \AND $\left\| \overline{\up}^k -  \overline{\up}^{k-1} \right\| > 0 $} \label{inalgoANMC:while}
		\STATE {Calculate $\vm d^k$ by solving Newton's equation $\nabla^2  Y\left(\overline{\up}^k\right)\vm d^k = -\nabla  Y\left(\overline{\up}^k\right)$.}
		\IF {"Calculation of $\vm d^k$ possible" \textbf{and} $-\nabla  Y\left(\overline{\up}^k\right)\vm d^k \geq \min \left( \change{\upvarphi}_1, \change{\upvarphi}_2 \left\| \vm d^k \right\|^q \right) 										\left\| \vm d^k \right\|^2$}
		\STATE {Set search direction $\vm s^k = \vm d^k$.}
		\ELSE
		\STATE {Set search direction $\vm s^k = -\nabla  Y\left(\overline{\vm p}^k\right)$.}
		\ENDIF
		\STATE {Determine step size with Armijo rule, i.e., search for largest $\sigma^k \in \left\{ \beta^0, \beta^1, \beta^2, \beta^3 \right\}$ \\
			~~~~such that: $Y\left(\overline{\up}^k +\sigma^k \vm s^k  \right) - Y\left(\overline{\up}^k\right) \leq \sigma^k \gamma \nabla  Y\left(\overline{\up}^k\right)^{\transpose} \vm s^k$, else set $\sigma^k = \beta^3$. } \label{inalgoANMC:inequal}
		\STATE {Set $\overline{\up}^{k+1}=\overline{\up}^k + \sigma^k \vm s^k$ and $k=k+1$.}
		\ENDWHILE \label{inalgoANMC:endwhile}
		\STATE {Calculate standard deviation $\sigma_Y = \sqrt{\frac{Y(1-Y)}{\Nmc}}$}\label{inalgoANMC:std}
		\IF {$\sigma_Y > \sigma_{Y,\mathrm{max}}$} \label{inalgoANMC:targetAcc}
		\WHILE {$\sigma_{Y'} > \sigma_{Y,\mathrm{max}}$ \textbf{and} $\left| Y\left(\overline{\up}^k\right) - Y'\left(\overline{\up}^k\right) \right| < \sigma_{Y,\mathrm{max}}$} \label{inalgoANMC:IncreaseRules}
		\STATE {Increase sample size $\Nmc^{\text{new}} = \Nmc + \text{inc} \, \Nmc^{\text{start}}$.}\label{inalgoANMC:increase}
		\STATE {Calculate $Y'\left(\overline{\up}^k\right)$ and $\sigma_{Y'}$ with $\Nmc^{\text{new}}$.}\label{inalgoANMC:recalc}
		\STATE {Set $\Nmc = \Nmc^{\text{new}}$.}
		\ENDWHILE
		\STATE{Set $Y\left(\overline{\up}^k\right) = Y'\left(\overline{\up}^k\right)$.} \label{inalgoANMC:update}
		\STATE{Go back to line \ref{inalgoANMC:while}.}\label{inalgoANMC:back2while}
		\ELSE
		\STATE {Stop with $\overline{\up}^{\star} = \overline{\up}^k$}\label{inalgoANMC:stop}
		\ENDIF
	\end{algorithmic}
	\label{algo:globAdaptNewton}
\end{algorithm}
A pseudo code for the adaptive Newton-\gls{mc} is given in Algorithm \ref{algo:globAdaptNewton}.
First, we need to define a target accuracy in form of a maximal standard deviation $\sigma_{Y,\mathrm{max}}$ for our terminal solution. Furthermore, we have to define the size of the initial MC sample $\Nmc^{\text{start}}$ and an incremental factor $\text{inc}>0$ such that
\begin{equation}
\Nmc^{\text{new}} = \Nmc^{\text{old}} + \text{inc} \, \Nmc^{\text{start}}.
\end{equation}
The sample size is increased until the target accuracy is reached (see line~\ref{inalgoANMC:targetAcc} in Algorithm~\ref{algo:globAdaptNewton}), and the standard globalized Newton method terminates because no further yield improvement can be obtained, i.e., the difference between $\overline{\up}^k$ and $\overline{\up}^{k-1}$ tends to zero (see line~\ref{inalgoANMC:while}). In line~\ref{inalgoANMC:IncreaseRules} we can see the rules for a sample size increment. This loop is activated, if the two previous mentioned conditions are fulfilled. 
Then, we increase the sample size stepwise (see line~\ref{inalgoANMC:increase}), re-evaluate the yield with the new size $Y'(\overline{\up}^k)$, and its new standard deviation $\sigma_{Y'}$ (see line~\ref{inalgoANMC:recalc}).
Note that in order to estimate $Y'(\overline{\up}^k)$ it is not necessary to evaluate $\Nmc^{\text{new}}$ new sample points. Only the $\text{inc} \, \Nmc^{\text{start}}$ additional points have to be evaluated and can then be fused with the $\Nmc^{\text{old}}$ old points to obtain the new yield estimator. 
This procedure is repeated until the new standard deviation $\sigma_{Y'}$ reaches the target accuracy (i.e., $\sigma_{Y'} \leq \sigma_{Y,\mathrm{max}}$) or the improvement of the yield is large enough (i.e., the difference between the actual yield $Y\left(\overline{\up}^k\right)$ and the yield with the increased sampling $Y'\left(\overline{\up}^k\right)$ is larger than the target accuracy $\sigma_{Y,\mathrm{max}}$). In that case we start a new iteration of the Newton algorithm, with updated yield and sample size (see line~\ref{inalgoANMC:back2while}). If the target accuracy is fulfilled after a regular Newton procedure (after line~\ref{inalgoANMC:endwhile}), the algorithm terminates (see line~\ref{inalgoANMC:stop}).

The parameters are chosen as for Algorithm~\ref{algo:globNewton}, additionally we set the maximal standard deviation, the starting sample size and the incremental factor as follows
\begin{equation}
\sigma_{Y,\mathrm{max}} = 0.01,~~ \Nmc^{\text{start}} = 100,~~ \text{inc}=1.
\end{equation}
Another difference in comparison to Algortihm~\ref{algo:globNewton} is, that we bound the number of Armijo backward steps. If the inequality in line \ref{inalgoANMC:inequal} is not fulfilled after three steps, we set $\sigma^k = \beta^3$ and proceed with the next iteration.

\section{Numerical results}\label{sec:NumericalResults}
We apply the methods for yield estimation and optimization discussed in the previous sections to a benchmark problem in the context of electromagnetic field simulation. In particular, we employ the model of a rectangular waveguide with a dielectric inset, similarly to the one used in \cite{loukrezis2019assessing}. This model is well suited for validation purposes, as a closed-form solution is available \cite{loukrezis2019adaptive}. \change{Additionally, it fulfills the assumption of a smooth input-to-output behaviour made in Section~\ref{subsubsec:SC}.} In the following, we first introduce the problem setting before numerical results for yield estimation as well as yield optimization are presented.

\subsection{Problem setting}
Starting from the time-harmonic Maxwell's equation on a computational domain $\dom\subset\mathbb R^3$, one can derive the curl-curl equation
\begin{equation}
\nabla \times \left(\mu^{-1} \nabla \times \vm E_{\freq} \right)  - \freq^2 \eps \vm E_{\freq} = 0 \quad \change{\text{in~}}\dom
\label{eq:E-field_strong_general}
\end{equation}
to be solved for the electric field phasor $\vm E_{\freq}$, where $\omega$ denotes the angular frequency,
$\mu=\mu_{\change{\mathrm r}} \mu_0 \in L^\infty(D)$ the dispersive complex magnetic permeability and $\change{\varepsilon}=\change{\varepsilon}_{\change{\mathrm r}}\change{\varepsilon}_0 \in L^\infty(D)$ the dispersive complex electric permittivity, with the vacuum permeability $\mu_0$ and the relative permeability $\mu_{\change{\mathrm r}}$, respectively vacuum and relative permittivity $\change{\varepsilon}_0$ and $\change{\varepsilon}_{\change{\mathrm r}}$.
Further we have assumed absence of charges and source currents. \change{Relating \eqref{eq:E-field_FEM} to the general problem \eqref{eq:Intro_KontMod} introduced in the beginning, we note that the angular frequency $\omega$ corresponds to the range parameter $\rp$. } 

\begin{figure}
	\centering
	\includegraphics[]{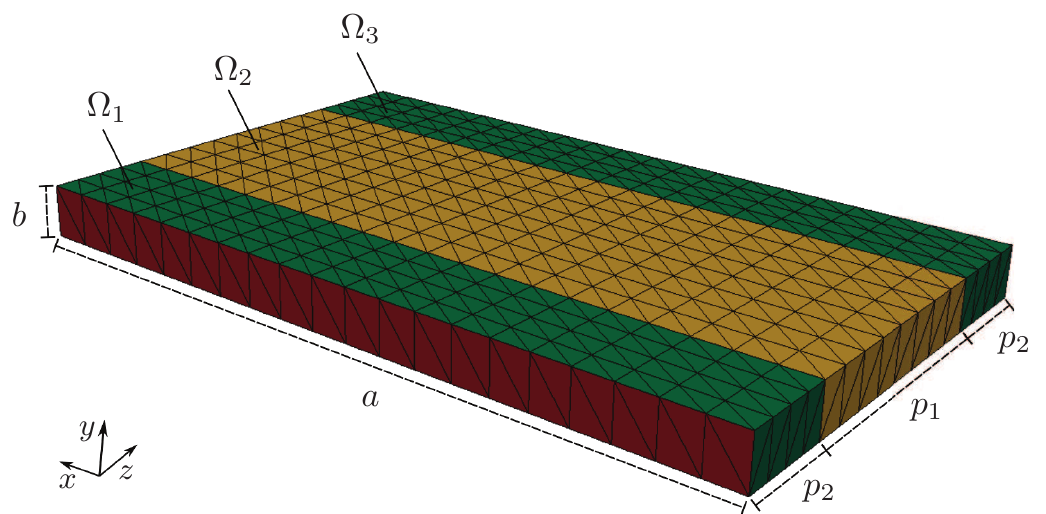}
	\caption{Finite element model of a rectangular waveguide with dielectric inset of length $p_1$. The waveguide is excited at the port $\Gamma_{\text{P1}}$ by an incident TE$_{10}$ wave.}
	\label{fig:waveguide}
\end{figure}

The boundary of the domain $D$ is split into three parts, i.e., $\partial D=\Gamma_{\text{PEC}}\cup\Gamma_{\text{P1}}\cup\Gamma_{\text{P2}}$, since we consider the model of an electric waveguide with two ports $\Gamma_{\text{P1}}, \Gamma_{\text{P2}}$ and assume \gls{pec} boundary conditions at the waveguide walls, i.e.,
\begin{equation}
\mathbf n \times \mathbf E_\omega = 0 \quad \text{on~}\Gamma_\text{PEC}.\label{eq:pec_bc}
\end{equation}
At the waveguide ports $\Gamma_{\text{P1}},\Gamma_{\text{P2}}$ we impose lowest order waveguide boundary conditions \cite[Chapter 8.5]{jin2015finite}
\begin{subequations}
	\noeqref{eq:1,eq:2}
	\begin{align}
	\vm n \times ( \nabla \times \vm E_{\freq})  - jk_{z10}(\vm n \times \vm E_{\freq})\times \vm n & = -2jk_{z10} \vm E_\freq^\text{inc} && \text{on } \Gamma_{\text{P1}}, \label{eq:1}\\
	\vm n \times (\nabla \times \vm E_{\freq}) - jk_{z10}(\vm n \times \vm E_{\freq})\times \vm n & = 0 && \text{on } \Gamma_{\text{P2}}, \label{eq:2}
	\end{align}
	\label{eq:wg_bc}
\end{subequations}
where $\vm n$ denotes the outer unit normal vector \change{and $j$ the imaginary unit}. The propagation constant $k_{z10}$ is given by $k_{z10}=\sqrt{\omega^2\mu_0\change{\varepsilon}_0-\frac{\pi}{a^2}}$, where, in turn, 
$a$ denotes the width of the waveguide, as depicted in Fig.~\ref{fig:waveguide}. According to~\cite{Jackson_1998aa}, the boundary conditions \eqref{eq:wg_bc} can be derived based on the assumption, that the rectangular waveguide is excited at $\Gamma_{\text{P1}}$  by an incident TE$_{10}$ wave 
\begin{equation}
\vm E_\freq^\text{inc}=E_0 \vm E^\text{TE}_{10}e^{-jk_{z10} z}\quad\text{with}~\vm E^\text{TE}_{10}:=\sin\left(\frac{\pi x}{a}\right)\mathbf e_y ,
\end{equation}
where $E_0$ refers to the amplitude of the incident wave and $\mathbf e_y$ denotes the unit vector in $y-$direction.   Additionally it is assumed that the waveguide dimensions are chosen s.t. only the TE$_{10}$ mode is propagating without attenuation,  that the ports are placed sufficiently far from any obstacles in the waveguide which might excite higher-order modes and that the homogeneous material at the ports $\Gamma_{\text{P1}}\cup\Gamma_{\text{P2}}$ fulfills $\epsr=\mur=1$. For further details on waveguide boundary conditions, we refer to \cite{jin2015finite}.

As \gls{qoi} we consider the fundamental scattering parameter (S-parameter) of the TE$_{10}$-mode on $\Gamma_{\text{P}1}$
\begin{equation}
S:= \frac{2
}{E_0 ab}\left( \vm E_\freq - \vm E_\freq^{\text{inc}}, \vm E^\text{TE}_{\text{10}} \right)_{\Gamma_{\text{P}1}},	\label{eq:S-Para_general}
\end{equation}
where we assumed \change{that $\Gamma_{\text{P}1}$ is located at $z=0$} for simplicity (without loss of generality). Note that the QoI \eqref{eq:S-Para_general} is, in this case, an affine-linear functional of $\vm E_{\freq}$.

\subsection{Weak formulation and discretization}
In order to solve the boundary value problem \eqref{eq:E-field_strong_general}-\eqref{eq:wg_bc} numerically by the \gls{fem}, we devise the corresponding weak formulation. Therefore, we build the inner products of \eqref{eq:E-field_strong_general} with test functions $\vm E'\in V$, where $V$ is to be determined, and integrate by parts \change{employing \cite[Theorem 3.31]{monk2003finite}} 
\begin{align}
\left( {\mur}^{-1} \nabla \times \vm E_{\freq}, \nabla \times \vm E' \right)_{\dom} & - \freq^2 \mu_0 \left( \eps \, \vm E_{\freq}, \vm E' \right)_{\dom}  + \left( \traceop[\mur^{-1} \nabla \times \vm E_\freq],\Traceop[\vm E'] \right)_{\partial D} = 0.
\label{eq:E-field_weak_wg2}
\end{align}
Note that we introduced the trace operators
\begin{align}
\traceop[\vm u]&:=\vm n \times \vm u|_{\partial D}\\
\Traceop[\vm u]&:=\bigl(\vm n\times \vm u|_{\partial D}\bigr)\times \vm n
\end{align}
for brevity of notation.
The boundary integral in \eqref{eq:E-field_weak_wg2} vanishes on $\Gamma_{\text{PEC}}$, since we impose \gls{pec} boundary conditions \eqref{eq:pec_bc} for the test functions $\vm E'$ as well. On $\Gamma_{\text{P1}}\cup\Gamma_{\text{P2}}$ we employ the boundary conditions \eqref{eq:wg_bc} and obtain the weak formulation: find $\vm E\in V$ s.t.
\begin{align}
\begin{split}
&\left( {\mur}^{-1} \nabla \times \vm E_{\freq}, \nabla \times \vm E' \right)_{\dom}  - \freq^2 \mu_0 \left( \eps \, \vm E_{\freq}, \vm E' \right)_{\dom}  + j k_{z10} \left(\Traceop[\vm E_\freq],\Traceop[\vm E'] \right)_{\Gamma_{\text{P1}}\cup\Gamma_{\text{P2}}} \\&= 2j k_{z10} \left(\vm E^\text{inc}_\freq,\Traceop[\vm E'] \right)_{\Gamma_{\text{P1}}}\quad \quad\forall \vm E' \in V.
\label{eq:E-field_weak_wg}
\end{split}
\end{align}
The appropriate function space $V$ is a subspace of 
\begin{equation}
H(\text{curl},\dom) := \left\lbrace  \vm u \in \left( L^2(\dom) \right)^3: (\nabla \times \vm u, \nabla \times \vm u)_{\dom} < \infty \right\rbrace,
\end{equation}
where, in turn, $\left( L^2(\dom) \right)^3$ denotes the complex vector function space of square integrable functions, i.e.,
\begin{equation}
\left( L^2(\dom) \right)^3 := \left\lbrace  \vm u: (\vm u, \vm u)_{\dom} < \infty \right\rbrace,
\end{equation}
cf. \cite{monk2003finite}. To account for the \gls{pec} boundary conditions \eqref{eq:pec_bc} and obtain a well-defined boundary integral in \eqref{eq:E-field_weak_wg}, $V$ is chosen as 
\begin{align}
V:= \left\lbrace \vm u \in H(\text{curl},\dom): ~\Traceop[\vm u]\big|_{\Gamma_{P1}} \hspace{-.2em}\in \left( L^2(\Gamma_{P1}) \right)^3 ~\wedge~ \Traceop[\vm u]\big|_{\Gamma_{P2}}\hspace{-.2em} \in \left( L^2(\Gamma_{P2}) \right)^3 ~\wedge~ \traceop[\vm u]\big|_{ \Gamma_{\text{PEC}} }=0 \right\rbrace. \label{eq:V}
\end{align}

In order to solve \eqref{eq:E-field_weak_wg} with \gls{fem}, we introduce a 
finite-dimensional function space $V_h \subset V$ and express the electric field as
\begin{equation}
\vm E_{\freq,h} = \sum_{j=1}^{\change{N_h}} e_{\freq,j} \, \vm N_j,
\label{eq:discreteE-field}
\end{equation}
where $e_{\omega,j} \in \CC$ are the degrees of freedom (DoF), $\change{N_h}$ is the number of DoFs and $\vm N_j \in V_h$ denotes second order, first kind Nédélec basis functions defined on a tetrahedral mesh of the domain $D$. For further details on the curl-conforming discretization, we refer to \cite{Nedelec_1980aa}.
The discrete solution $\mathbf e_\omega=[e_{\freq,1}, \ldots, e_{\freq,\change{N_h}}]^{\transpose}$ is then obtained by solving the linear system 
\begin{equation}
\underbrace{ \left( \vm K   - \freq^2 \vm M^{\change{\varepsilon}} + jk_{z10}\vm M^{\text{port}} \right) }_{\vm A_{\freq}}
\underbrace{\vm e_{\freq}}_{\vm e_{\freq}}
= \underbrace{\vm f(\vm e^{\text{inc}})}_{\vm f_\freq} 
\label{eq:E-field_FEM}
\end{equation}
where $\vm A_{\freq} \in \CC^{\change{N_h} \times \change{N_h}}$ is the system matrix and $\vm f_{\freq} \in \CC^{\change{N_h}}$ is the discretized right-hand side. The stiffness matrix $\vm K$, the mass-matrix $\vm M^\change{\varepsilon}$, the matrix $\vm M^\text{port}$ and the right-hand side $\vm f_\freq$ in the above expression are given by
\begin{align}
\begin{aligned}[c]
\vm K_{ij}&= (\mu_{\change{\mathrm r}}^{-1} \nabla \times \vm N_j, \nabla \times \vm N_i)_D,\\
\vm M^\text{port}_{ij} &=  \left(\Traceop[\vm N_j],\Traceop[\vm N_i] \right)_{\Gamma_{\text{P1}}\cup\Gamma_{\text{P2}}},
\end{aligned}\hspace{8em}
\begin{aligned}
\vm M^{\change{\varepsilon}}_{ij}&= \mu_0 (\change{\varepsilon} \vm N_j, \vm N_i)_D,\\
[\vm f_\freq]_i &= 2j k_{z10} \left(\vm E^\text{inc}_\freq,\Traceop[\vm N_i] \right)_{\Gamma_{\text{P1}}}.
\end{aligned}
\end{align}
The S-parameter can then be obtained from the discete counterpart of \eqref{eq:S-Para_general} 
\begin{equation}
S_h(\freq) = \bigl(\vm q_{\freq} , \vm e_{\freq}- \vm e_{\freq}^\text{inc}\bigr)_{{\mathbb C}^{N_h}}.
\label{eq:S-Para_FEM}
\end{equation}

As discussed in the previous sections, we then introduce a parameter vector $\up \in \Xi \subset \mathbb R^M$ to account for variations in the design parameters, which, in this case, might represent changes in the domain $D$ or in the material parameters $\change{\varepsilon}, \mu$. Hence, we obtain the parametrized discrete system
\begin{subequations}
	\begin{align}
	\vm A_{\freq} (\up) \, \vm e_{\freq}(\up) &= \vm f_{\freq}  \label{eq:E-field_FEM_p}, \\
	S_h(\up,\freq) &= \bigl(\vm q_{\freq}, \vm e_{\freq}(\up)\bigr)_{{\mathbb C}^{N_h}}\change{-\bigl(\vm q_{\freq},\vm e_{\freq}^\text{inc}\bigr)_{{\mathbb C}^{N_h}}}.  \label{eq:S-Para_FEM_p}
	\end{align}
\end{subequations}
\change{We note that the S-parameter is an affine-linear functional in this case, while we only considered linear functionals in Section~\ref{subsubsec:SC} for brevity of notation. However, the method can be straightforwardly adapted to address the constant offset $\bigl(\vm q_{\freq} , \vm e_{\freq}^\text{inc}\bigr)_{{\mathbb C}^{N_h}}$ such that the adjoint-based error indicators remain valid.}

We proceed with a few details on the implementation of the numerical model. To assemble the linear system \eqref{eq:E-field_FEM_p}, we employ the \gls{fe} library \textsc{FEniCS} \cite{AlnaesBlechta2015a}. As \textsc{FEniCS 2017.2.0} does not support complex numbers, we assemble real and imaginary parts of the matrices separately. We then use \textsc{numpy} to impose the \gls{pec} boundary condition \eqref{eq:pec_bc} and \textsc{scipy} to solve the resulting linear system of equations with a sparse-LU decomposition. Employing the readily available LU decomposition, the corresponding dual solution $\mathbf z_\omega(\up)$ can then also be obtained with negligible costs, since the dual problem 
\begin{equation}
\vm A^{\star}_{\freq} (\up) \mathbf z_\freq(\up) = \vm q_\freq,
\end{equation}
can again be solved by forward-backward substitution.

\subsection{Numerical results}
We consider twelve uncertain parameters 
\begin{equation}
\up = \left[ p_1, \dots , p_{12} \right]^{\transpose}.
\end{equation}
Two of them are geometrical parameters given in mm (length of the dielectrical inlay $p_1$ and length of the vacuum offset $p_2$) and ten are material parameters with effect on the relative permittivity $\change{\varepsilon}_{\change{\mathrm r}}|_{\Omega_2}$ and permeability $\mu_{\change{\mathrm r}}|_{\Omega_2}$ on the dielectrical inlay
\begin{align*}
\change{\varepsilon}_{\change{\mathrm r}}|_{\Omega_2} &= p_5 + \left(p_3-p_5\right)\left(1+j\freq p_6 \tau\right)^{-1} + 
\left(p_4-p_5\right)\left(1+j\freq p_7 \tau\right)^{-1}, \\
\mu_{\change{\mathrm r}}|_{\Omega_2} &=  p_{10} + \left(p_8-p_{10}\right)\left(1+j\freq p_{11} \tau\right)^{-1} + 
\left(p_9-p_{10}\right)\left(1+j\freq p_{12} \tau\right)^{-1},
\end{align*}
where
\begin{align*}
\freq &= 2 \pi f, \\
\freq_0 &= 2 \pi \left(20 \cdot 10^9\,\mathrm{Hz}\right),\\
\tau &= \frac{1}{\freq_0}
\end{align*}
with frequency $f$ (in Hertz).
In order to consider the influence of the number of uncertain parameters, tests with four uncertain parameters are also performed. For this purpose we consider a modified parameter vector
\begin{equation}
\up^{(4)} = \left[ p_1, p_2, p_{13},  p_{14} \right]^{\transpose},
\end{equation}
where $p_1$ and $p_2$ are the geometrical parameters from above, and $ p_{13}$ and $p_{14}$ are material parameters with the following effect on relative permeability and permittivity
\begin{align*}
\change{\varepsilon}_{\change{\mathrm r}}^{(4)}|_{\Omega_2} &= 1+p_{13} + \left(1-p_{13}\right)\left(1+j\freq \left( 2 \pi 5 \cdot 10^9 \right)^{-1}\right)^{-1}, \\
\mu_{\change{\mathrm r}}^{(4)}|_{\Omega_2} &= 1 + p_{14} + \left(2-p_{14}\right)\left(1+j\freq \left( 1.1\cdot 2\pi 20 \cdot 10^9 \right)^{-1}\right)^{-1}.
\end{align*}
For yield optimization we set the starting point $\overline{\up}_0$ for twelve parameters to
\begin{equation}
\overline{\up}_0 = [9,5,2,0.5,1,1,1.1,2.5,1,1,1,2]^{\transpose}.
\end{equation}
The estimation tests are done for a reference value $\overline{\up}_{\text{e}}$ close to one optimal solution
\begin{equation}
\overline{\up}_{\text{e}} = [8.6,3.8,2,0.5,0.7,0.6,1.4,2.8,1.7,0.8,0.3,1.4]^{\transpose}.
\end{equation}
For the tests with four parameters we set the starting points to
\begin{align*}
\overline{\up}_0^{(4)} &= [\change{9},5,1,1]^{\transpose}, \\
\overline{\up}_{\text{e}}^{(4)} &= [10.36, 4.76, 0.58, 0.64]^{\transpose}.
\end{align*}
The standard deviation is set to $\sigma = 0.7^2$\,mm for geometrical, and $\sigma = 0.3^2$ for material parameters. In order to avoid unphysical values, instead of a normal distribution we use a truncated normal distribution for the \gls{mc} sample generation. We truncate with an offset $t$ of $\pm 3$\,mm and $\pm 0.3$ for the geometrical and material parameters, respectively.
The performance feature specifications are
\begin{equation}
\left| S(\up,\freq)\right|  \stackrel{!}{\leq} -24 \, \text{dB} \ \ \forall \freq \in \IRPo_{\freq} = \left[ 2\pi f_1, 2\pi f_2 \right] = \left[ 2 \pi 6.5, 2 \pi 7.5 \right] \text{ in GHz.}
\end{equation}
\change{Related to the setup of the performance feature specifications in~\eqref{eq:pfs} this means  $c = -24$\,dB and $\QoI(\up,\rp) = \left| S(\up,\freq)\right|$ with the frequency $\freq$ as range parameter.}
We consider eleven equidistant frequency points $\freq_j \in \IRPo_{\freq_j}$ in the frequency range.
The reference solution for yield estimation is
\begin{equation}
Y_{\text{Ref}}^{(12)}(\overline{\up}) = 74.60\,\%,
\end{equation}
\change{for twelve uncertain parameters and 
	\begin{equation}
	Y_{\text{Ref}}^{(4)}(\overline{\up}) = 95.44\,\%,
	\end{equation}
	for four uncertain parameters.
	Both reference solutions have been} calculated with a closed-form solution of the E-field formulation and standard \gls{mc} method with $\Nmc=2,500$, which is the smallest sample size fulfilling $\sigma_{Y,\mathrm{max}} = 0.01$ for all sizes of the yield, according to~\eqref{eq:StdMC}.

\subsubsection{Quality of the gradient}

As mentioned in section~\ref{sec:YieldOptimization} there is a difference between differentiating or discretizing first. Furthermore, for the sample generation we use a truncated normal distribution instead of a normal distribution. Thus, the gradient we use for optimization deviates from the exact gradient, which can be thought of as an inexact Newton method~\cite{Dembo_1982}, 
with approximations in the root-finding problem itself, i.e., here the gradient~\eqref{eq:Grad}, and the Jacobian which is in our case the Hessian~\eqref{eq:Hess}.

\begin{figure}
	\centering
	\includegraphics[]{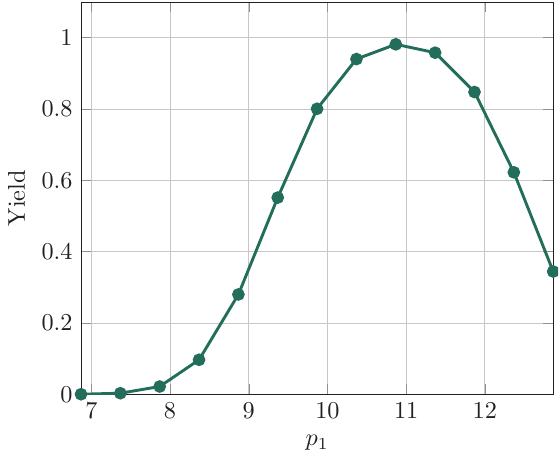}
	\hspace{2em}
	\includegraphics{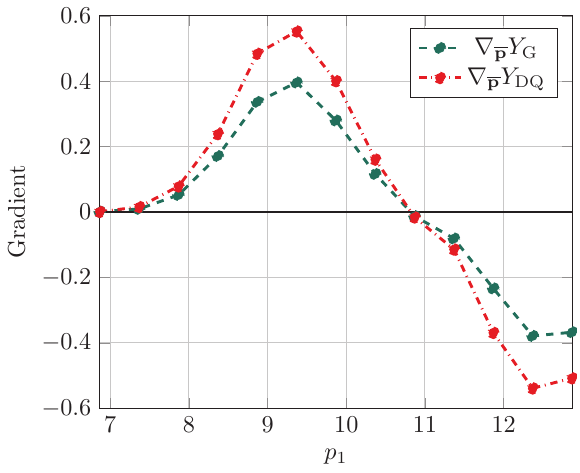}
	\caption{Comparision of the gradients $\nabla_{\overline{\up}}Y_{\text{G}}$ and $\nabla_{\overline{\up}}Y_{\text{DQ}}$ for $\Nmc=10^6$ and finite difference step size $\change{\delta}=10^{-3}$ for truncated normal distribution.}
	\label{fig:CompareGrads}
\end{figure}

\begin{figure}
	\centering
	\includegraphics[]{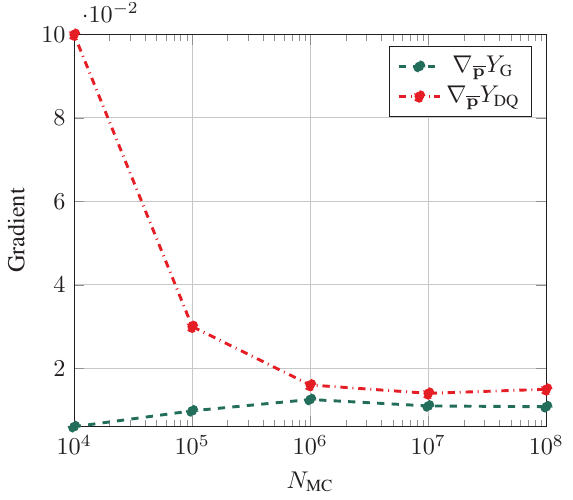}
	\caption{Convergence of the gradients $\nabla_{\overline{\up}}Y_{\text{G}}$ and $\nabla_{\overline{\up}}Y_{\text{DQ}}$ for increasing $\Nmc$. Calculated in $p=10.8685$\,mm and with finite difference step size $\change{\delta}=10^{-3}$.}
	\label{fig:ConvGrads}
\end{figure}

To ensure that the yield is optimal at the end and no further improvement is possible, an extension can be added to the optimization algorithm. At the optimal solution, the gradient can be calculated with a finite difference quotient $\nabla_{\overline{\up}}Y_{\text{DQ}}$. 
The gradient from~\eqref{eq:Grad} will be denoted $\nabla_{\overline{\up}}Y_{\text{G}}$, to avoid any confusion.
We consider the difference between the two gradients and expect it to be smaller than a constant $\change \eta$
\begin{equation}
\left| \nabla_{\overline{\up}}Y_{\text{DQ}}(\overline{\up}) - \nabla_{\overline{\up}}Y_{\text{G}}(\overline{\up}) \right| \leq \change \eta.
\label{eq:QualityOfGrad}
\end{equation}
Figure~\ref{fig:CompareGrads} compares the two gradients $\nabla_{\overline{\up}}Y_{\text{G}}$ and $\nabla_{\overline{\up}}Y_{\text{DQ}}$ for the waveguide example where the only uncertain design parameter is the length of the inlay \change{$p_1$}. On the left, we see the yield over the parameter \change{$p_1$}, on the right we see the graphs of the gradients over the parameter \change{$p_1$}. For this calculation we set the sample size to $\Nmc=10^6$ and the step size in the difference quotient to $\change{\delta}=10^{-3}$. The two gradients show a similar behaviour, especially near the optimum the gradients agree well. 
Figure~\ref{fig:ConvGrads} shows how the two gradients approach each other for large $\Nmc$. 
Thus, if~\eqref{eq:QualityOfGrad} is not fulfilled the number of sample points to calculate the gradients can be increased until~\eqref{eq:QualityOfGrad} is fulfilled or an upper bound for $\Nmc$ is reached. In the former case, the applied gradient $\nabla_{\overline{\up}}Y_{\text{G}}$ is accurate and the optimal solution reliable. In the latter case, a further improvement of the yield would still be possible due to the limited gradient accuracy. In this case the yield optimization could be continued with the gradient $\nabla_{\overline{\up}}Y_{\text{DQ}}$. However, this would require additional computational effort, especially for a large number of uncertain parameters. The optimal solution can also be used as a starting point for an alternative optimization procedure.

\subsubsection{Yield estimation}

We proceed by comparing the proposed hybrid approach with standard \gls{mc} and a surrogate-based \change{\gls{mc}} approach without hybridization. The surrogate model is constructed based on sparse-grid interpolation as explained in Section~\ref{subsubsec:SC}. In order to achieve a high accuracy in the $L^\infty$-norm, we employ, in this work, uniform weight functions $w_m$ in the ranges given by the nominal parameter values $\overline{\up}$ $\pm$ truncation offset $t$. The comparison is based on both the computational effort and the accuracy. 
For the accuracy we use the relative error between the reference solution and the solution of the considered method, i.e., for the hybrid approach
\begin{equation}
\err_{\text{H}} = \frac{\left|Y_{\text{Ref}}-Y_{\text{H}}\right|}{Y_{\text{Ref}}},
\label{eq:relerr}
\end{equation}
$\err_{\text{SC}}$, $\err_{\text{MC}}$ for SC and \gls{mc}, respectively. 
We measure the computational effort with the number of \change{high fidelity} evaluations (i.e., matrix factorizations in \gls{fem}). Here we have different levels of \change{high fidelity} evaluations due to mesh refinement within the proposed hybrid approach. 
We start with a mesh size $h$, and if necessary divide it by two. The difference in the computational effort for each refinement level depends on the model structure and the solver used. Assuming an optimal solver with an effort which is linear in the number of degrees of freedom, the effort increases by a factor of four in the case of a 2D problem and by a factor of eight in the case of a 3D problem. 
Since in our case the E-field is constant in y-direction, the grid is only refined in x- and z-direction.
Thus, the computational effort of a method is measured through 
\begin{equation}
\text{eff} = \# \text{HF}_{h} + 4 \, \# \text{HF}_{h/2} + 16 \, \# \text{HF}_{h/4} 
\label{eq:comp.eff.}
\end{equation}
which adds up the number of \change{high fidelity} evaluations on the different levels, each multiplied by the factors mentioned above. 

The standard approach to carry out yield estimation, with the same accuracy as with the proposed hybrid approach, would be a \gls{mc} analysis with the finest mesh used within the hybrid approach, referred to as $\text{MC}_{\text{fine}}$. If the mesh refinement strategy is additionally applied, the method is denoted as $\text{MC}_{\text{refine}}$. In order to build the surrogate model both for SC and for the hybrid approach, we use the first grid with mesh size $h$ without further refinement to evaluate the model at the Leja nodes. 
\begin{table}
	\centering
	\caption{Comparison of different yield estimation approaches for twelve uncertain parameters.}
	\begin{filecontents*}{data_table1.csv}
approach;Leja;HF h surr.;HF h MC;HF h2 MC;HF h4 MC;eff;err
MC$_{\textrm{fine}}$;-;0;0;0;22705;363,280;0.0000
MC$_{\textrm{refine}}$;-;0;22705;25;6;22,901;0.0000
H;90;990;4812;25;6;5,998;0.0000
SC;90;990;0;0;0;990;6.2235
SC;1,600;17600;0;0;0;17,600;0.4290
\end{filecontents*}
\pgfplotstabletypeset[
	col sep=semicolon,
	every row 2 column 0/.style={postproc cell content/.style={@cell content=\textbf{##1}}},
	every row 2 column 6/.style={postproc cell content/.style={@cell content=\textbf{##1}}},
	every row 2 column 7/.style={postproc cell content/.style={@cell content=\textbf{##1}}},
	every head row/.style={%
		before row={\toprule},
		after row=\midrule
	},
	every last row/.style={after row=\bottomrule},
	columns/approach/.style={column name=approach, string type, column type=c},
	columns/Leja/.style={column name=$\#$ Leja, string type, column type=r},
	columns/HF h surr./.style={column name=$\#$ HF$_h$ surr., column type=r},
	columns/HF h MC/.style={column name=$\#$ HF$_h$ MC, column type=r},
	columns/HF h2 MC/.style={column name=$\#$ HF$_{h/2}$ MC, column type=r},
	columns/HF h4 MC/.style={column name=$\#$ HF$_{h/4}$ MC, column type=r},
	columns/eff/.style={column name=eff, string type, column type=r},
	columns/err/.style={column name=err $(\%)$, fixed, fixed zerofill, precision=4, column type=r},
	multicolumn names
	]{data_table1.csv}
	{\\ Different methods: MC, SC and hybrid (H). $\#$ Leja indicates the number of Leja nodes for one frequency point $\freq_j$, $\#$ HF the number of high fidelity evaluations to build the surrogate model (surr.), to evaluate (critical) MC samples (MC) with indicated refinement, eff the measurement for computational effort according to~\eqref{eq:comp.eff.} and err the relative error according to~\eqref{eq:relerr}.}
	\label{tab:CompareMethodsEst_wg12D}
\end{table}
In Table~\ref{tab:CompareMethodsEst_wg12D} we see the results of the comparison. We consider two versions of SC, each with different accuracy (number of Leja nodes). The surrogate model used for the hybrid approach is the same as for SC with 90 Leja nodes. For each approach we use the same \gls{mc} sample as for the reference solution. With the hybrid approach and \gls{mc} we achieve the same result as with the closed-form reference solution. Out of these three, the hybrid method requires the least computational effort. Compared to MC$_{\text{refine}}$, we can save $73\,\%$ computing time, compared to MC$_{\text{fine}}$ even $98\,\%$. Comparing the hybrid and MC$_{\text{refine}}$ approach, we observe that most of the \gls{mc} sample points are evaluated on the coarsest \gls{fe} grid. Only for a few points, a refinement of the grid to $h/2$ (25 sample points) or $h/4$ (6 sample points) is necessary.
Using the same surrogate model as for the hybrid approach, pure SC has much less computational effort with eff $=990$, but the error is larger than $6\,\%$. Increasing the number of Leja nodes to $1,600$ results in three times higher computational effort compared to the hybrid approach, with an error still larger than $0.4\,\%$.

Table~\ref{tab:CompareMethodsEst_wg4D} shows the results for the same waveguide with only four uncertain parameters. The statement remains unchanged. However, the influence of the number of parameters can be seen. With four uncertain parameters also with SC we can reduce the error to zero, with only about two and a half times the computational effort compared to the hybrid method. The higher the number of uncertain parameters, the more gain can be expected from the hybrid approach compared to a SC method.
Compared to MC$_{\text{refine}}$, with the hybrid approach, we can save almost $98\,\%$ computing time, compared to MC$_{\text{fine}}$ even $99.8\,\%$. This means that the advantage of the hybrid approach over \gls{mc} increases as the number of parameters decreases. Nevertheless, we know by construction, that even for high numbers of uncertain parameters, the hybrid method can never become worse than \gls{mc}, excluding the computational effort to build the surrogate model (which could scale poorly for a high-dimensional problem) and evaluate the error indicator.

\begin{table}
	\caption{Comparison of different yield estimation approaches for four uncertain parameters.}
	\centering
	\begin{filecontents*}{data_table2.csv}
approach;Leja;HF h surr.;HF h MC;HF h2 MC;HF h4 MC;eff;err
MC$_{\textrm{fine}}$;-;0;0;0;26360;421,760;0.0000
MC$_{\textrm{refine}}$;-;0;26360;5;1;26,396;0.0000
H;30;330;165;5;1;531;0.0000
SC;30;330;0;0;0;330;0.1257
SC;120;1320;0;0;0;1,320;0.0000
\end{filecontents*}
\pgfplotstabletypeset[
	col sep=semicolon,
	every row 2 column 0/.style={postproc cell content/.style={@cell content=\textbf{##1}}},
	every row 2 column 6/.style={postproc cell content/.style={@cell content=\textbf{##1}}},
	every row 2 column 7/.style={postproc cell content/.style={@cell content=\textbf{##1}}},
	every head row/.style={%
		before row={\toprule},
		after row=\midrule
	},
	every last row/.style={after row=\bottomrule},
	columns/approach/.style={column name=approach, string type, column type=c},
	columns/Leja/.style={column name=$\#$ Leja, string type, column type=r},
	columns/HF h surr./.style={column name=$\#$ HF$_h$ surr., column type=r},
	columns/HF h MC/.style={column name=$\#$ HF$_h$ MC, column type=r},
	columns/HF h2 MC/.style={column name=$\#$ HF$_{h/2}$ MC, column type=r},
	columns/HF h4 MC/.style={column name=$\#$ HF$_{h/4}$ MC, column type=r},
	columns/eff/.style={column name=eff, string type, column type=r},
	columns/err/.style={column name=err $(\%)$, fixed, fixed zerofill, precision=4, column type=r},
	multicolumn names
	]{data_table2.csv}
	{\\ Different methods: MC, SC and hybrid (H). $\#$ Leja indicates the number of Leja nodes for one frequency point $\freq_j$, $\#$ HF the number of high fidelity evaluations to build the surrogate model (surr.), to evaluate (critical) MC samples (MC) with indicated refinement, eff the measurement for computational effort according to~\eqref{eq:comp.eff.} and err the relative error according to~\eqref{eq:relerr}.}
	\label{tab:CompareMethodsEst_wg4D}
\end{table}

\subsubsection{Yield optimization}

We compare the proposed adaptive Newton-\gls{mc} from Algorithm~\ref{algo:globAdaptNewton} with the standard Newton method from Algorithm~\ref{algo:globNewton}, both with the same limited number of Armijo backsteps and the presented hybrid approach for the yield estimation. In both cases we set the target accuracy to $\sigma_{Y,\mathrm{max}} = 0.01$. The adaptive approach starts with 100 sample points and increases this number adaptively until optimality and a target accuracy are achieved. In the non-adaptive approach, we specify a fixed sample size so that the target accuracy is guaranteed at all times during the optimization process. This fixed sample size is $\Nmc = 2,500$.
In Table~\ref{tab:CompareMethodsOpt_wg4-12D} we see the results for tests with twelve or four uncertain parameters. The number of iterations of single yield estimations within the optimization process, the computational effort (eff) and the optimal yield value ($Y^{\star}$) are given. 
Note that, during the optimization process the surrogate model is only built once, for the starting point. 
Accepting higher computational effort, the surrogate model can also be recalculated in each iteration step for the current solution, or built at the beginning in a larger interval than $\overline{\up}_0 \pm t$.

With twelve uncertain parameters, we started with a yield of $15\,\%$. The adaptive and the non-adaptive approach lead to different local optima with similar yield values. Both take a bit more than 30 iterations. 
On average, two and a half yield estimations are performed per iteration using the adaptive approach. This is due to multiple evaluations by Armijo backsteps. The non-adaptive approach has only 1.2 estimations per iteration. 
This can be explained by the fact that the adaptive approach performs several Newton optimizations with different sample sizes one after the other. Shortly before a Newton procedure is terminated, there is usually no further improvement, which is why Armijo backsteps increase and so does the number of yield estimations. This is the case every time before the sample size is increased in the adaptive algorithm. In the non-adaptive approach, this behaviour occurs only once at the end. Potential for improvement in the adaptive approach lies in further reducing the number of yield evaluations through smoother transitions from one sample size to the other. Nevertheless, the adaptive approach reduces the computational effort by a factor of two compared to standard Newton, see column $\text{eff}$ in Table~\ref{tab:CompareMethodsOpt_wg4-12D}. 
In tests with only four uncertain parameters, the computing effort was even reduced to $10\,\%$. In this case, the adaptive approach resulted in significantly fewer iterations. The ratio between iterations and yield estimations remains unchanged.
\begin{table}
	\caption{Comparison of adaptive and nonadaptive Newton method for yield optimization.}
	\centering
	\begin{filecontents*}{data_table3.csv}
est;Leja;UQ para;opt;It;YE;eff;Y opt
H;90;12;adapt. Newton-MC;33;86;376,073;74.84
H;90;12;Newton;37;42;682,745;78.20
H;30;4;adapt. Newton-MC;12;27;13,716;95.44
H;30;4;Newton;30;34;138,158;97.92
\end{filecontents*}
\pgfplotstabletypeset[
	col sep=semicolon,
	every row 3 column 1/.style={after row=\hline},
	every head row/.style={%
		before row={\toprule},
		after row=\midrule
	},
	every last row/.style={after row=\bottomrule},
	columns/est/.style={column name=estimation, string type, column type=c},
	columns/Leja/.style={column name=$\#$ Leja, column type=c},
	columns/UQ para/.style={column name=$\#$ param., column type=c},
	columns/opt/.style={column name=optimization, string type, column type=c},
	columns/It/.style={column name=$\#$ It, column type=r},
	columns/YE/.style={column name=$\#$ YE, column type=r},
	columns/eff/.style={column name=eff, string type, column type=r},
	columns/Y opt/.style={column name=$Y^{\star}$ $(\%)$, fixed, fixed zerofill, precision=2, column type=r},
	multicolumn names
	]{data_table3.csv}
	{\\ Comparision of adaptive and nonadaptive Newton's method with twelve and four uncertain parameters: $\#$ Leja indicates the number of Leja nodes for one frequency point $\freq_j$, $\#$ param. the number of uncertain parameters, optimization the method used, $\#$ It the number of iterations, $\#$ YE the number of yield estimations, eff the computational effort and $Y^{\star}$ the optimal yield value.}
	\label{tab:CompareMethodsOpt_wg4-12D}
\end{table}

For the case with four uncertain parameters we also draw a comparison to standard procedures.
Standard procedure means in this case, combining a standard \change{\gls{mc}} analysis for the yield estimation with a standard Newton method for the optimization.
On the coarsest grid $(h)$, $816,816$ evaluations with FEM were necessary to optimize the yield, i.e., $\text{eff} = 816,816$. Thus, with the proposed adaptive Newton-MC, a saving of $98.3\,\%$ in computing effort could already be achieved compared to the standard procedure mentioned above.
However, in order to achieve the same accuracy as with the proposed method, the finest grid $(h/4)$ has to be used for all sample points. We assume that the number of function evaluations does not change significantly due to the grid refinement.
This can be motivated by the fact, that a similar number of iterations, yield estimations and function evaluations were needed for calculation with the closed-\change{form} solution as for the \gls{fe} model with coarser grid. Under this assumption we got an effort factor of $\text{eff} \approx 13 \cdot 10^6$. Thus the saving of computational effort is even $99.9\,\%$.

For twelve uncertain parameters, in Table~\ref{tab:OptProcess_wg12D} we see how many \gls{mc} samples have been used in which iteration. For most of the iterations a low number of \gls{mc} samples is sufcient, only in the last iterations we need to expend more computational effort in order to guarantee the pre-defined target accuracy.
\begin{table}
	\caption{Progress of yield optimization with adaptive Newton-MC.}
	\centering
	\begin{filecontents*}{data_table4.csv}
iteration; MC samples
0-12;100
13-14;200
15-18;300
19;500
20-22;600
23;900
24-30;1,000
31;1,800
32-33;1,900
\end{filecontents*}
\pgfplotstabletypeset[
	col sep=semicolon,
	every head row/.style={
		before row={\toprule},
		after row=\midrule
	},
	every last row/.style={after row=\bottomrule},
	columns/iteration/.style={column name=iteration, string type, column type=c},
	columns/MC samples/.style={column name=$\Nmc$, string type, column type=c},
	]{data_table4.csv}
	{\\ Progress of yield optimization with adaptive Newton-MC for twelve uncertain parameters. Number of MC sampels for each iteration of the optimization algorithm.}
	\label{tab:OptProcess_wg12D}
\end{table}

\section{Conclusion}\label{sec:Conclusion}

In this paper we proposed \change{an adaptive} method for yield estimation and optimization. For yield estimation we developed a hybrid approach combining reliability and accuracy of a high fidelity Monte Carlo (MC) analysis and the efficiency of surrogate based techniques such as stochastic collocation. In case the accuracy of the surrogate model is not sufficient, sample points are re-evaluated employing the high fidelity finite element (\gls{fe}) model. Mesh refinement is applied if the accuracy of the \gls{fe} model itself is too low. This guarantees error control while only a very small subset of the MC sample is evaluated based on expensive high fidelity evaluations.
Adjoint error indicators were applied to estimate the errors of the surrogate model and the \gls{fe} model. For yield optimization we proposed an adaptive Newton-MC method, based on a globalized Newton method. 
During the optimization process, numerous yield estimations are performed. In order to control the MC error and at the same time save computational effort, we adaptively increase the number of MC sample points used during the optimization. Thus, the adaptive Newton-MC in combination with the hybrid approach allows us to control the \gls{fe} error, the MC error and the surrogate error. At the same time it is much more efficient than conventional MC approaches with a standard Newton method. Numerical tests on a dielectrical waveguide confirm the benefits of the presented method. 
Future research will deal with the transitions in the adaptive Newton-MC when the MC sample size is increased.
Furthermore, although we already use a hierarchical model for Monte Carlo analysis within the optimization, we plan to explore a combination of this with a multilevel Monte Carlo approach~\cite{Giles_2015aa}.


\acknowledgements

This work is supported by the Excellence Initiative of the German Federal and State
Governments and the Graduate School of Computational Engineering at Technische
Universität Darmstadt. N. Georg and U. R\"omer acknowledge funding by the Deutsche Forschungsgemeinschaft (DFG, German Research Foundation) RO4937/1-1. \change{The authors would like to thank two anonymous reviewers for their helpful comments that improved the manuscript.}

\appendix

\section{Abbreviations}\label{app:abbrev}

\change{
	\begin{tabular}{ll}
		FE & finite element \\
		FEM & finite element method \\
		MC & Monte Carlo \\
		PEC & perfect electric conductor\\
		QoI & quantity of interest \\
		SC & stochastic collocation \\
		TE & transverse electric
	\end{tabular}
}

\section{Symbols}\label{app:symbols}

\change{
	\begin{longtable}[l]{ll}
		English & \\
		$\vm A_{\up,\rp}$  & general system matrix		\\
		$\vm A^{\star}_{\up,\rp}$ & Hermitian transpose of $\vm A_{\up,\rp}$ \\
		$\vm A_{\omega}$  & system matrix of waveguide model\\
		$a$ & waveguide width\\
		$b$ & waveguide height\\
		$c$ & upper bound to define the performance feature specifications \\
		$D$ & computational domain \\
		$\vm E'$ & test functions\\
		$\vm E_\omega$ & electric field phasor \\
		$\vm E^\text{inc}_\omega$ & incident wave phasor (excitation)\\
		$\vm E_{\omega,h}$ & finite element approximation of $\vm E_\omega$  \\
		$E_0$ & amplitude of incident wave\\
		$\vm E^\text{TE}_{10}$ & fundamental transverse electric mode\\
		$\vm e_{\omega}$ & vector of degrees of freedom \\
		$e_{\omega,j}$ & degree of freedom \\
		$\vm e_y$ & unit vector in $y$-direction\\
		$\text{eff} $ & computational effort for yield estimation or optimization\\
		$\text{err} $ & error of the yield estimator compared to the reference solution\\
		$f$ & frequency\\
		$\vm f_{\rp}$ & general discrete right-hand side	\\
		$\vm f_{\omega}$ & discrete right-hand side	of waveguide model\\
		$g_{\rp}$ & forcing term \\
		$H(\text{curl}, D)$ & complex function space of square integrable functions with square integrable curl\\
		$h$ & mesh size for FEM \\
		$\text{inc}$ & incremental factor for the adaptive Newton method \\
		$j$ & imaginary unit \\
		$\vm K $ & stiffness matrix\\
		$k_{z10}$ & propagation constant\\
		$L_{\up,\rp}$ & parametric differential operator \\
		$L^{\star}_{\up,\rp}$ & adjoint operator of $L_{\up,\rp}$\\
		$L^2(D)$ & complex function space of square integrable functions on $D$\\
		$L^\infty(D)$ & complex function space of essentially bounded functions on $D$\\
		$\vm M^\varepsilon$ &  mass matrix\\
		$\vm M^\text{port}$ & system-matrix contribution stemming from port boundary conditions\\
		$\vm N_j$ & second order, first kind N\'ed\'elec basis functions \\
		$N_{h}$ & number of degrees of freedom \\
		$\Nmc$ & size of the Monte Carlo sample \\
		$\Nmc^{\text{start}}$ & initial size of the MC sample for adaptive yield optimization\\
		$\Nmc^{\text{new}}$ & updated size of the MC sample in adaptive yield optimization\\
		$\Nmc^{\text{old}}$ & old size of the MC sample in adaptive yield optimization\\
		$N_{\up}$ & number of uncertain input parameters \\
		$\Nsc$ & number of interpolation nodes \\
		$N_{\SD}$ & number of accepted sample points \\
		$\vm n$ & outer unit normal vector\\
		$\up$ & vector of uncertain input parameters $\left( \up = [p_1,...,p_{n_{\up}}] \right)$\\
		$\up_i$ & realization of the input parameter vector $\up$ \\
		$\up^{(i)}$ & interpolation nodes\\
		$\upj$ & uncertain input parameter \\
		$p_1, p_2$ & length of the inlay, length of the offset of the waveguide \\
		$p_3,\dots, p_{12}$ & material parameters of the waveguide \\
		$\overline{\up}$ & mean value of the uncertain input parameter vector $\up$ \\
		$\overline{\up}_0$ & mean value of the starting point for yield optimization \\
		$\overline{\up}_{\text{e}}$ & mean value of the considered point for yield estimation \\
		$\overline{\up}_{\SD}$ & mean value of the accepted sample points	\\
		$\hat{\overline{\up}}_{\SD}$ & MC approximation of the mean value of the accepted sample points	\\
		$\pdf$ & probability density function \\
		$Q$ & quantity of interest \\
		$Q_h$  & finite element approximation of quantity of interest \\
		$q$ & parameter for angular condition in Newton method  \\
		$(q_r,\cdot)_D$ & general linear functional defining  the quantity of interest	\\
		$(\vm q_r, \cdot)_{\mathbb C^{n_h}}$ &	general discrete linear functional defining the discrete quantity of interest \\
		$(\vm q_\omega, \cdot)_{\mathbb C^{n_h}}$ &	discrete linear functional of the waveguide model\\
		$\rp$ & range parameter \\
		$\rp_j$ & range parameter point \\
		$S$ & scattering parameter of the fundamental transverse electric mode on $\Gamma_\text{P1}$\\
		$S_h$ & finite element approximation of $S$\\
		$s$ & safety factor \\
		$s^k$ & search direction in the $k$-th step of the Newton method \\
		$\IRP$ & range to define the performance feature specifications \\
		$\dIRP$ & discretized range to define the performance feature specifications \\
		$u_{\rp}$ & solution of the model problem \\
		$\vm u_{\rp}$  & discrete primal solution \\
		$V$ & function space defined in equation~\eqref{eq:V}\\
		$V_h$ & finite-dimensional subspace of $V$\\
		$w$ & weight function \\
		$Y$ & yield \\
		$Y_{\text{MC}}$ & MC estimator of the yield \\
		$Y_{\text{Ref}}$ & Reference value of the yield for numerical tests \\
		$\nabla_{\overline{\up}} Y_{\text{DQ}}$ & gradient of the yield according to differential quotient\\
		$\nabla_{\overline{\up}} Y_{\text{G}}$ & analytical gradient of the yield according to~\eqref{eq:Grad} \\
		$\vm z_r$ &	general discrete dual solution	\\
		$\vm z_\omega$ &	discrete dual solution of waveguide model\\
		Greek & \\
		$\alpha_i$ & coefficients for stochastic collocation \\
		$\beta$ & parameter for Armijo rule in Newton method \\
		$\Gamma_\text{P1}, \Gamma_{\text{P2}}$ & waveguide ports\\
		$\Gamma_\text{PEC}$ & waveguide walls\\
		$\gamma$ & parameter for Armijo step size in Newton method \\
		$\delta$ & step size for differential quotient\\
		$\epsilon_{\text{fe}}$ & finite element error \\
		$\epsilon_{\text{sc}}$ & stochastic collocation error \\
		$\varepsilon_{0}, \varepsilon_{\text{r}}$ & vacuum and relative permittivity \\
		$\mu_0, \mu_{\text{r}}$ & vacuum and relative permeability \\
		$\Xi$ & image space of uncertain parameters of waveguide model\\
		$\traceop, \Traceop$ & tangential trace operators\\
		$\Sigma$ & covariance matrix of the uncertain input parameter vector $\up$ \\
		$\Sigma_{\SD}$ & covariance matrix of the accepted sample points	\\
		$\hat{\Sigma}_{\SD}$ & MC approximation of the covariance matrix of the accepted sample points	\\
		$\sigma^k$ & step size in the $k$-th step of the Newton method \\
		$\sigma_Y$ & standard deviation of the yield estimator \\
		$\sigma_{Y,\text{max}}$ & upper bound for the standard deviation of the yield estimator \\
		$\tau$ &  relaxation time \\
		$\Phi_i$ & global polynomial basis functions \\
		$\upvarphi_1, \upvarphi_2$ & parameters for angular condition in Newton method \\
		$\freq$ & angular frequency \\
		$\Omega_1, \Omega_3$ & domain of the vacuum of the waveguide \\
		$\Omega_2$ & domain of the dielectrical inlay of the waveguide \\
		$\SD$ & safe domain\\
		Other symbols & \\
		$\tilde{X}$  & stochastic collocation approximation of a function $X$ \\
		$\mathcal{I}^1_{\epsilon}$ & trusted interval (with estimated FE and SC error)\\	
		$\mathcal{I}^2_{\epsilon}$ & trusted interval (with FE error)\\
		$\# \text{HF}_h$ & number of high fidelity (FE) evaluations with grid refinement $h$
	\end{longtable}
}














\end{document}